\documentclass{article}

\usepackage{arxiv}

\usepackage[utf8]{inputenc} 
\usepackage[T1]{fontenc}    
\usepackage{hyperref}       
\usepackage{url}            
\usepackage{booktabs}       
\usepackage{amsfonts}       
\usepackage{nicefrac}       
\usepackage{microtype}      
\usepackage{lipsum}		
\usepackage{graphicx}
\usepackage{subcaption}
\usepackage{enumitem}

\usepackage{siunitx}
\usepackage[english]{babel}
\usepackage{amsmath,amssymb,amstext, bm}
\usepackage{color}
\usepackage[backend=biber, style=nature,isbn=false,url=false]{biblatex}
\DeclareSourcemap{
	\maps[datatype=bibtex]{
		\map{
			\step[fieldset=abstract, null]
		}
	}
}
\AtEveryBibitem{\clearfield{month}}
\AtEveryCitekey{\clearfield{month}}

\title{Sensitivity of principal components to system changes in the presence of non-stationarity}

\author{ Henrik M. Bette \\
	\texttt{henrik.bette@uni-due.de} \\
	\and
	Michael Schreckenberg \\
	\texttt{michael.schreckenberg@uni-due.de} \\
	\and
	Thomas Guhr \\
	\texttt{thomas.guhr@uni-due.de} \\
	\and
	Fakultät für Physik, University of Duisburg-Essen, Duisburg, Germany \\
}

\renewcommand{\shorttitle}{PCA sensitivity to system changes with non-stationarity}

\hypersetup{
	pdftitle={Sensitivity of principal components to system changes in the presence of non-stationarity},
	pdfsubject={},
	pdfauthor={Henrik M.~Bette, Thomas~Guhr},
	pdfkeywords={},
}

\begin{document}
	\maketitle
	
	\begin{abstract}
		Non-stationarity affects the sensitivity of change detection in correlated systems described by sets of measurable variables. We study this by projecting onto different principal components. Non-stationarity is modeled as multiple normal states that exist in the system even before a change occurs. The studied changes occur in mean values, standard deviations or correlations of the variables. Monte Carlo simulations are performed to test the sensitivity for change detection with and without knowledge about the non-stationarity for different system dimensions and numbers of normal states. A comparison clearly shows that the knowledge about the non-stationarity of the system greatly improves change detection sensitivity for all principal components. This improvement is largest for those components that already provide the greatest possibility for change detection in the stationary case. We illustrate our results with an example using real traffic flow data, in which we detect a weekend and a bank holiday start as anomalies.
	\end{abstract}

	\keywords{change detection \and non-stationarity \and nonlinear dynamics \and stochastic processes \and correlations \and principal component analysis}

\section{Introduction}

The detection of changes, novelty, failures or faults is a crucial task in numerous real world systems, such as industrial monitoring, energy generation, IT- and road-traffic, sensor networks, image processing and many more \cite{Pimentel2014, Kerner2020, Newhart2019, Maleki2019}. Early or even preemptive detection can help operations, avoid down-times and reduce costs in general. Techniques for such analysis can be roughly categorized into three groups: model-based methods, knowledge-based methods and data-driven methods \cite{Sanchez-Fernandez2018}. While a lot of research is done in all directions, the first two groups require extensive knowledge about the monitored system and are therefore challenging and time-consuming. This has made the data-driven methods especially interesting for researchers in all fields \cite{Kazemi2020}.

Principal Component Analysis (PCA) \cite{Denis2021, Pimentel2014, Newhart2019} is one of those data-driven methods. It is commonly used as either a direct detection method or as a dimensionality reducing pre-processing. Here, data is projected into the space of the principal components, which are the eigenvectors of the covariance or correlation matrix for a system with multiple measured variables. Whether covariances or correlations are used depends on the system studied. The eigenvectors associated with the large eigenvalues are referred to as the major components. Without a change present, these carry most of the information on the behavior of the system. A projection onto the subspace of the major components includes the largest part of the variance of a data set. The other eigenvectors associated with small eigenvalues are referred to as minor components. For change detection one usually calibrates a normal parameter set of these projections. New data is then projected and compared to the normal space using, for example, Hotelling $T^2$-statistics or the sum of squared prediction error together with a threshold \cite{Bakdi2017, Pozo2018}. It has been applied to change detection in many different systems, including wind turbines \cite{Vidal2018}, wastewater treatment plants \cite{Sanchez-Fernandez2018}, healthcare institutions \cite{Harrou2015} and traffic \cite{Li2019}. An important question is which of the principal components to look at. For the use-case of dimensionality reduction one generally employs a projection onto the major components \cite{Denis2021}. However, the residual subspace, i.e. the minor components, is often most useful for outlier and change detection \cite{Pimentel2014, Dutta2007, Kuncheva2014}. Shyu et. al. \cite{Shyu2003} proposed a method combining the components relating to the largest and the smallest eigenvalues for novelty detection. Recently, Tveten \cite{Tveten2019} studied the sensitivity of different principal components in a more structured and complete way confirming the high sensitivity of minor components to changes.

Many variants of PCA exist. They were developed to tackle different specific problems. Some better known ones include, but are not limited to, static and dynamic robust PCA \cite{Vaswani2018, Ahmadi2021}, kernel PCA \cite{Rahim2021}, dynamic kernel PCA \cite{Bounoua2021}, moving window PCA \cite{Haimi2016}, recursive PCA \cite{Elshenawy2018} and incremental PCA \cite{Kazemi2020}. The latter three tackle the important problem of non-stationarity. Many studied systems are subject to external influences or internal processes, that cause varying conditions. These lead to changes in the data, which do not represent novelty or failure. We refer the reader to the review \cite{Ketelaere2017} for an introduction to PCA on time-dependent data. A performance comparison is found in \cite{Rato2016}. The problem of non-stationarity is also of on-going interest for other novelty detection methods \cite{Shang2015, Canonaco2020, BuenoDeMesquita2021, Chen2019, Zimroz2014, Huang2022}. 

Non-stationarity exists in different forms. First, a system can slowly evolve due to influences such as climate \cite{BuenoDeMesquita2021} or economic situation \cite{Munnix2012}. On the other side, external conditions can vary on shorter time-scales causing non-stationarity without an evolutionary trend \cite{Zimroz2014, WangShanshan2020, Bette2022}. Some systems also exhibit periodic non-stationarity \cite{Chen2019}. A second distinction can be made in the type of non-stationarity: changes can occur in the absolute values of measured variables, their individual distribution or the relation between different variables. In simple terms these three categories represent novelty as changes in mean values, standard deviations or correlation structure respectively. Of course, this does not capture all possibilities. For example, distribution changes can also effect higher moments only. Furthermore, relations can be non-linear. However, this simplification already provided valuable insights in the sensitivity of principal components in stationary systems \cite{Tveten2019}. We study the importance of non-stationarity accordingly.

Of those three categories non-stationarity in the correlation structure is especially important for PCA (or any other method depending directly on the covariance or correlation matrix). The purpose of PCA is to find new coordinates that already incorporate the linear correlations in the system. If they change, i.e. the eigenvectors of the correlation matrix change their structure, the projections into PCA space are substantially different. Such a non-stationarity in the correlation matrix has been thoroughly investigated, for example, in financial markets \cite{Munnix2012, Heckens2020a}, road traffic \cite{WangShanshan2020} and wind turbine data \cite{Bette2022}. We assume that such a distinction into well-separated individual states is most likely to happen for a difference in correlation structure. At a certain point internal or external causes lead to a behavioral change. Changes in mean or standard deviation of variables might also exhibit such distinct states, but are more likely than a correlation structure to change continually. Therefore, our analysis will focus on correlation changes, but we will nevertheless study all types of change.

We aim with this study for a structured and comprehensive analysis of the sensitivity of principal components in the presence of non-stationarity, while trying to keep it simple enough that mechanisms can be easily understood. The non-stationarity is modeled as multiple possible normal states of the system. We study all three types of possible changes, but do not mix them. This means, if the change occurring at a certain point in time is of a certain type (mean value, standard deviation, correlation structure), the states will also differ in that same type. We base the study framework on the one used by Tveten \cite{Tveten2019} and extend it to incorporate non-stationarity. This facilitates comparison with the stationary case. We analyze the sensitivity of the principal components under the assumption that we know about the non-stationarity and compare it to the results obtained, if we did not know about the non-stationary behavior. Thereby, we can study how important knowledge about such a state-wise non-stationarity is for novelty detection with PCA.

The main focus of our study is a general analysis of the influence of non-stationarity on change detection with PCA as well as the presentation of a method to account for this non-stationarity. Additionally, we illustrate our simulation results with an example using real traffic flow data. Thereby, we show how the proposed methodology can be applied to measured data. In traffic, the non-stationarity is common knowledge: traffic volumes are not at all constant during a day. Furthermore, traffic is very different on workdays as compared to weekends or bank holidays. These well known facts make traffic a good example to understand the application of the method. We first classify the non-stationarity during single work days. Then, we define weekends and the onset of a bank holiday as a change and try to detect these with and without taking the classified non-stationarity into account.

We describe the idea of the experiment with a metaphor in section \ref{sec:experimentIdea}. In section \ref{sec:theory} we introduce the problem and notations and set up the simulation framework in a general form. Section \ref{sec:results} contains the results of our simulation studies divided into subsections for the different change types. In these we also present the detailed simulation set up for each case. We apply the method to real data by working out the traffic example in section \ref{sec:trafficexample}. A summary of results and findings is found in section \ref{sec:conclusion}.

\section{Analysis Concept}
\label{sec:experimentIdea}

To facilitate easy understanding of the concept of our analysis before going into detail, we outline the simulation experiment with a simple example. Imagine a lamp with a light bulb that is burning with a certain color. Let us say it emits blue light. At a certain point in time $t=\tau$ the light bulb changes its color slightly and we study from which viewing angle we can best detect this change. As a measure we take a color chart and estimate the difference $h$ in points between the color before and after the change. Then we repeat this experiment time and again with different bulbs, different colors and different color changes while keeping track of the change detectability from each viewing angle. This is an easily imagined example for the stationary simulation experiment.

Now, let us assume that before $t=\tau$ the light bulb burns with $S$ different colors. At random intervals it changes its color, but all colors are equally likely over time. This represents non-stationarity in the way we are going to analyze. At time $t=\tau$ a change occurs once again. However, the light bulb can be in any state $s \in {1,...,S}$, i.e. burn with one of the possible colors. If we know of the non-stationarity and have a criterion to determine which color the lamp should have, we detect the change compared to the realized color $s^{*}$. Once again we use our color scale and determine a point score $h_1$ from different viewing angles. However, we want to analyze how much improvement in the detectability of that change is caused by our knowledge about this non-stationarity. Therefore, we also calculate a detectability without it: We compare the color after the change occurs with the average color before the state. So, if our lamp would normally burn with either blue or red, we compare to a violet light. Again we take the color chart and note difference $h_2$ in points between changed color and average color before change from each angle.

To measure the increase in detectability with knowledge of the non-stationarity we have to compare the two measurements $h_1$ and $h_2$. The first one is easy enough, because it should produce the same results as the stationary case if our knowledge about the original state, i.e. color, of the light bulb is perfect. When measuring without knowledge about the non-stationarity we have to take into account that even without a change we will detect deviations from our assumed normal (the average state): At any given time the light will be red or blue, but we will compare it to violet and observe a difference $h_{s,\mathrm{norm}}$ between state $s$ and the average state. This means that changes at $t=\tau$ are detectable only if they differ stronger from the violet than our normal states (blue and red) do. We have to correct our measurement of difference by subtracting the maximum difference we measure without a change present: $h_2^{\mathrm{corr}} = h_2 - \max\limits_{s}( h_{s,\mathrm{norm}})$. As in the stationary case we perform multiple Monte-Carlo simulations. The importance of knowledge about non-stationarity for change detection is then measured as $h_1 - h_2^{\mathrm{corr}}$.

\section{Theory and Setup}
\label{sec:theory}

In our analysis, the light bulb from the example above is a $D$-dimensional system with $n$ observations $\bm{x}_t \in \mathbb{R}^D$ at times $t=1,\dots,n$. We do not look at the time series themselves, but rather describe the system with mean values, standard deviations and Pearson-correlations. This description is complete only for Gaussian systems. We assume this to be the case for our study. At the time $t=\tau$ a change occurs in the system, that can either influence mean, standard deviation or correlation structure of the observed variables. This is analogous to the change in color of the light bulb. For the stationary case the sensitivity to change in principal components was already analyzed by Tveten \cite{Tveten2019}. We extend it by the inclusion of non-stationarity.  We assume that the system has $S$ possible states and at every time is in one $s \in \{1, \dots, S\}$ of them. For each of these states $s$ we have a vector of means $\bm{\mu}_0^{(s)}$, a vector of standard deviations $\bm{\sigma}_0^{(s)}$ and a correlation matrix $C_0^{(s)}$. The index $0$ indicates that these values refer to the system before the change at $t=\tau$ occurs. In our example with the light bulb these three values together describe the color of the state. For further analysis, we standardize each state separately to mean zero and standard deviation one and denote standardization with a hat. This results in $\bm{\hat{\mu}}_0^{(s)} = \bm{0}$ and $\bm{\hat{\sigma}}_0^{(s)} = \bm{1}$. Without knowledge about the $S$ different states, two ways of measurement present themselves. One either measures just one set of parameters for the time span $t=1,\dots,\tau$ as was assumed in \cite{Tveten2019} or one measures the parameters in epochs of a length $\tau_{\mathrm{ep}}<\tau$ and takes the averages. The second option offers the advantage that new data, in which one wants to test for change, is most likely also only available for a time span $\tau_{\mathrm{new}}<\tau$. The comparability of new and old data is therefore increased if epoch length and time span of the new data coincide, i.e. $\tau_{\mathrm{ep}}\approx\tau_{\mathrm{new}}$. Correlations, for example, might be largely positive on a large time scale due to a global trend in values, but show more structure when measured on shorter time spans. Therefore we define the average state (analogous to the violet light of the light bulb example) as averages over epochs. Without loss of generality, we assume that each state $s$ occurs for the same amount of time before the change and define the parameters of the average state as element-wise averages,
\begin{equation}
	\bm{\overline{\mu}}_0 = \sum_{s=1}^{S} \bm{\mu}_0^{(s)}, ~~~
	\bm{\overline{\sigma}}_0^2 = \sum_{s=1}^{S} \bm{\sigma}_0^{2(s)}, ~~~
	\overline{C}_0 = \sum_{s=1}^{S} C_0^{(s)}.
\end{equation}
We average the variance instead of the standard deviation. The average over the correlation matrix is also meant element-wise and is according to the centroid calculation in real data clustering in e.g. \cite{Munnix2012, Heckens2020a, WangShanshan2020, Bette2022}. Standardization with the mean value and standard deviation of this average state will be denoted with a tilde: $\bm{\tilde{\overline{\mu}}}_0 = \bm{0}$ and $\bm{\tilde{\overline{\sigma}}}_0 = \bm{1}$.

When the change happens at time $t=\tau$, it will affect the values we currently measure. In the light bulb example this would be the slight change of color. We denote these new values with an index $1$ as $\bm{\mu}_1$, $\bm{\sigma}_1$ and $C_1$. Next, we need to decide how the data after the change should be standardized. In reality this questions is, how the new (or possibly live) data of a system should be standardized before comparing it to the usual behavior. This is independent of whether or not non-stationarity has been accounted for. One can either standardize with the known pre-change values of the usual behavior or with the newly measured ones. If one measures only a single new data point, only the first option is feasible. Without loss of generality, we normalize with pre-change values. However, the correlation matrix after the change will not be a well-defined correlation matrix. Especially the entries on its diagonal will not be one. This does not present a problem for our analysis.

Including non-stationarity in the way of multiple pre-change states, we need to know what normal state the system would have been in after the change at $t=\tau$ to make the right comparison. We will denote this state with $s^{*}$. In reality this means that after having found multiple normal states, we need one criterion (or possibly several criteria) marking the states. For the current, theoretical analysis we assume that we simply know the state $s^{*}$ in the epoch after the change. We can then standardize with the pre-change state parameters to get $\hat{\bm{\mu}}_1$, $\hat{\bm{\sigma}}_1$ and $\hat{C}_1$. The hat denotes standardization with state parameters. To compare the sensitivity of principal components with knowledge of non-stationarity to the case, in which we do not know about the different states, we will also calculate $\tilde{\bm{\mu}}_1$, $\tilde{\bm{\sigma}}_1$ and $\tilde{C}_1$, where the tilde denotes standardization with the average state parameters.

To analyze the sensitivity of principal components, we study the projections of the system onto the eigenvectors before and after the change. This is analogous to looking at the light bulb from different angles. For a state $s$ we calculate the eigenvalues $\lambda_j^{(s)}$ and normalized eigenvectors $\bm{v}_j^{(s)}$, $j=1, \dots,D$ of the correlation matrix $C_0^{(s)}$. Ordering is assumed from largest to smallest eigenvalue, i.e. $\lambda_1^{(s)} \geq \dots \geq \lambda_D^{(s)}$. Note that the eigenvectors are equivalent to the principal components. The projection of a data point $\bm{x}_t$ onto the $j$-th component is then calculated as $y_{j,t} = \bm{v}_j^{(s)T} \bm{x}_t$. Furthermore, we want to compare projections onto these components for epochs rather than a single data point. If we know the vector of means $\bm{\mu}$ and the correlation matrix $C$ of said epoch, we calculate the mean $\mu_j^{\prime}$ and standard deviation $\sigma_j^{\prime}$ of the projection onto the $j$-th eigencomponent by
\begin{equation}
	\label{eq:projection}
	\mu_j^{\prime} = \bm{v}_j^{(s)T} \bm{\mu} ~~~ \text{and} ~~~
	\sigma_j^{\prime} = \sqrt{\bm{v}_j^{(s)T} C \bm{v}_j^{(s)}}.
\end{equation}
We have dropped the state identification $(s)$ on the left side. This is possible, because projections are only ever needed into the average system and the system of $s^{*}$ so that no confusion with other states is possible. This allows us to keep the identifier, if it is actually the mean vector or the correlation matrix of the state, which is being projected. For example, the mean of projection onto component $j$ of the of the original state mean $\bm{\mu}_0^{(s)}$ would read $\mu_{j,0}^{\prime(s)}$, whereas the projection of the changed state means $\bm{\mu}_1$ would simply be $\mu_{j,1}^{\prime}$. Then also the overline of an average state would translate into an overline on the projected notation. If standardization notation is necessary it will also translate. For comparison without knowledge about the non-stationarity projections into the eigensystem $\overline{\lambda}_j$ and $\bm{\overline{v}}_j, ~j=1,\dots,D$ of the average state are also necessary. We denote these in the same way with a double prime,
\begin{align}
	\mu_j^{\prime\prime} &= \bm{\overline{v}}_j^{T} \bm{\mu} ~~~ \text{and}\\
	\sigma_j^{\prime\prime} &= \sqrt{\bm{\overline{v}}_j^{T} C \bm{\overline{v}}_j}.
\end{align} \\
We now have all projections into the eigenvector system, i.e. onto the principal components, we need. In short, we have for every component $j$ the projections of \begin{itemize}
	\item the original state into the correct state $\mu_{j,0}^{\prime(s^{*})}$, $\sigma_{j,0}^{\prime(s^{*})}$,
	\item the changed state into the correct state $\mu_{j,1}^{\prime}$, $\sigma_{j,1}^{\prime}$,
	\item the original state into the average state $\mu_{j,0}^{\prime \prime(s^{*})}$, $\sigma_{j,0}^{\prime \prime(s^{*})}$ and
	\item the changed state into the average state $\mu_{j,1}^{\prime \prime}$, $\sigma_{j,1}^{\prime \prime}$.
\end{itemize}
Having established the mean and standard deviation projections, we must now compare them. Therefore, we define the sensitivity for change as the Hellinger distance \cite{Hellinger1909} between the marginal distributions before and after the change. In the light bulb example this is the difference in points on the color chart. In general, the squared Hellinger distance between two probability measures $P$ and $Q$ on the Lebesque-measurable space $X$ with probability densities $p(x)$ and $q(x)$ is defined as
\begin{equation}
	\label{eq:HellDistDef}
	H^2(P,Q) = \frac{1}{2} \int\limits_{X} \left(\sqrt{p(x)} - \sqrt{q(x)}\right)^2 \mathrm{d}x= 1-\int\limits_{X} \sqrt{p(x)q(x)} \mathrm{d}x.
\end{equation}
For our case, it is easily calculated for two normal distributions $\mathcal{N}(\mu_1, \sigma_1^2)$ and $\mathcal{N}(\mu_2, \sigma_2^2)$ as
\begin{equation}
	\label{eq:HellDist}
	H^2(\mu_1, \sigma_1, \mu_2, \sigma_2) = 1- \sqrt{\frac{\sigma_1 \sigma_2}{\sigma_1^2 + \sigma_2^2}} \exp \left\lbrace - \frac{1}{4} \frac{(\mu_1-\mu_2)^2}{\sigma_1^2 + \sigma_2^2} \right\rbrace .
\end{equation}
For this theoretical study it is more feasible than a data distribution comparison such as Hotelling $T^2$-statistics. For the Hellinger distance to be applicable - under the assumption of normal distributions - we do not need the actual distributions. This allows us to simulate mean values, standard deviations and correlation matrices without having to generate the underlying data.

In Eq. \eqref{eq:HellDist} we have already adopted a notation that is easy to use for our application by writing the Hellinger distance as a function of the means and standard deviations of the normal distributions to compare. The projected $\mu$ and $\sigma$ are calculated as described above and fully describe the comparison. For example, the Hellinger distance between the pre-change projection of the state $s^{*}$ onto itself and the projection of the changed state into that same system with standardization with state pre-change parameters would read
\begin{equation}
	H(\hat{\mu}_{j,0}^{\prime(s^{*})}, \hat{\sigma}_{j,0}^{\prime(s^{*})}, \hat{\mu}_{j,1}^{\prime}, \hat{\sigma}_{j,1}^{\prime}) = H_j(\hat{\mu}_{0}^{\prime(s^{*})}, \hat{\sigma}_{0}^{\prime(s^{*})}, \hat{\mu}_{1}^{\prime}, \hat{\sigma}_{1}^{\prime}).
\end{equation}
For the right side we have simply taken the index $j$, which must always be equal for all arguments in $H$, and moved it from the arguments to the function for better readability.

We carry out different Monte Carlo simulations for changes in the correlation structure, the mean and the standard deviation. Each time we will assume that the other factors stay constant and the change type is the same for the occurring change at $t=\tau$ and in between the states. The detailed scheme for the simulation will always be given in the sections dealing with the results as these differ for the different change types. In general we will simulate normal states, calculate the average state and then simulate a change at $t=\tau$. We will always perform this for a multitude of normal states and for each of these for a multitude of change scenarios. The after-change data will be standardized and projected into the corresponding state eigensystem as well as the average eigensystem, Hellinger distances are calculated respectively. We then take the expectation value $\overline{H}_j$ over the Monte Carlo runs.

While in real data some fluctuations are always expected, which lead to non-zero Hellinger distances even without a change occurring, this is true for both the state and the average case and can therefore be neglected in a comparison. Remembering our light bulb example however, we have to correct the result without knowledge of the non-stationarity. In a non-stationary system with states there is an additional non-zero part of the Hellinger distance, if we do not know about the states. This is because the system would be in a state $s^{*}$, which is unequal to the average state even without change, but projected into the average eigensystem and compared to the average state. This would be additional noise in our change detection, that only occurs when not knowing about the non-stationarity. In our example this was the blue and red light already being different from the average violet color without a change. We therefore calculate the projections of the different states $s=1,\dots,S$ into the eigensystem of the average state. Of course, to do this, we need to standardize the state variables with the pre-change average parameters. We then calculate the Hellinger distances this would cause even without a change. The maximum of these distances is the threshold distance up to which we have to assume that the system behaves normally. We can only detect a real change at $t=\tau$ if it causes a Hellinger distance larger than the maximum one the states themselves cause. This maximum is subtracted from the Hellinger distance between the system with a change and the average state as only the difference between these two measures the detectability of a change in a non-stationary system without knowledge about this non-stationarity. As detectability can never be less than zero, we set negative values to zero. The subtraction is done inside the Monte Carlo runs. This ensures that the actual noise created by the state average comparison is taken as it can vary strongly depending on how different the original normal states are. We denote this corrected Hellinger distance with $\mathcal{H}_j$ .

\section{Results}
\label{sec:results}

\begin{figure}
	\centering
	\includegraphics[width=0.7\textwidth]{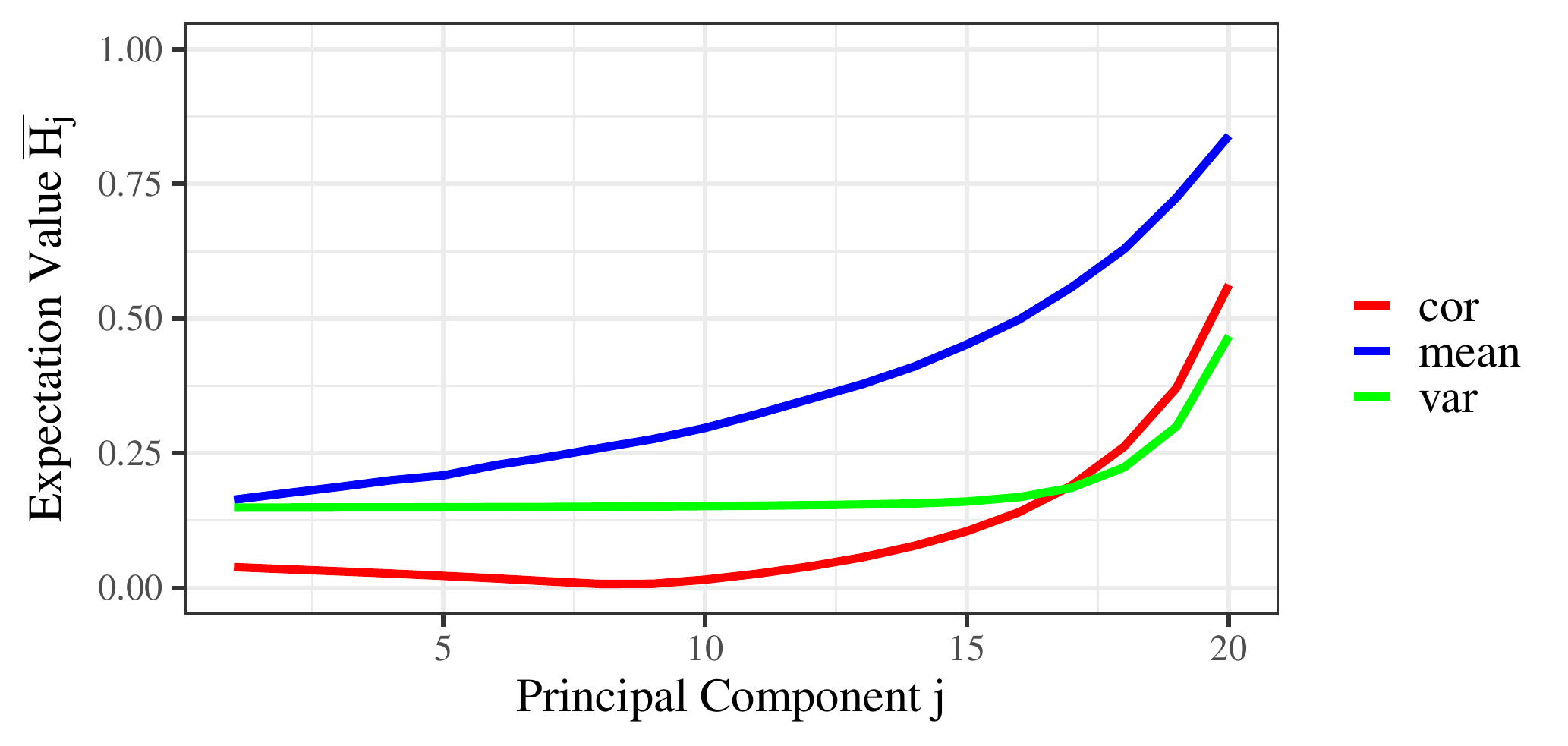}
	\caption{Monte Carlo estimates for the sensitivity to changes of the different eigenvectors in a stationary system. The sensitivities are the Hellinger distances between the distributions of data projections onto the eigenvectors before and after the change. Different types of changes are shown as different colors.}
	\label{fig:oneState}
\end{figure}

The focus of our analysis lies in the difference between knowing and not knowing about the non-stationarity of the system. We present our results separately for the different types of change. This way it is easier to describe the simulation procedure including necessary standardization. We have tested our implementation to produce the same results as given by Tveten \cite{Tveten2019} for the stationary case, i.e. if we assume the number of states $S$ to be one. The results are shown in Fig. \ref{fig:oneState}. It shows the sensitivity of the different eigenvectors to changes in the stationary system. As described in sec. \ref{sec:theory}, the sensitivity is given by the Hellinger distance between the distributions of data projections onto eigenvectors before and after the change. A higher value of the expectation value for the Hellinger distance on the y-axis therefore indicates a higher sensitivity to changes. Clearly, the minor components, i.e. the eigenvectors with small eigenvalues, display a higher sensitivity to change. This type of plot is used frequently in the following analysis including non-stationary. The interpretation of the values on the y-axis stays the same. Here, we show results for the three different types of changes in one plot. The simulation procedures used for each different change type are described in the following sections.

We present the simulation and its results for correlation structure, mean value and standard deviation  in Sec. \ref{sec:corresults}, \ref{sec:meanresults} and \ref{sec:sdresults}, respectively.

\subsection{Change in correlation structure}
\label{sec:corresults}

We explore the sensitivity of principal components to a change in the correlation structure of the simulated variables. We think that this is the prominent use-case for PCA as changes in mean and standard deviation can also be detected by comparing their values directly. In contrast to the other scenarios, we do not have to worry about standardization issues as mean and standard deviation always stay the same. This is so, because we assume that the change does not influence these parameters and the normal states only differ in correlation structure. Hence, the vector of means $\bm{\mu}_0^{(s)}$ and the vector of standard deviations $\bm{\sigma}_0^{(s)}$ are equal for all states. We will therefore refrain from using the standardization notation in this case for the sake of readability.

To obtain results for different combinations of dimension $D$, number of normal states $S$ and change sparsity $K$, we perform Monte Carlo simulations with various change scenarios. Thereby we get the estimate $\overline{H}_j$ as an average over all simulation runs. We simulate two scenarios, which differ in the method to create the different states $s$ that exist before the change. For one scenario they are random and unrelated to each other. For the second one we draw one random correlation matrix and obtain the other $S-1$ states by changing the first state in the same way that change is introduced at $t=\tau$ later on.

In the case of unrelated random states, our simulation follows the steps:
\begin{enumerate}
	\item Draw $S$ random correlation matrices $C_0^{(s)}, ~ s=1,\dots,S$ of dimension $D$ using the method described in \cite{Joe2006}.
	\item Calculate the element-wise average correlation matrix $\overline{C}_0$ before the change.
	\item Calculate the Hellinger distances $H_j(\overline{\mu}_0^{\prime\prime}, \overline{\sigma}_0^{\prime\prime},\mu_0^{\prime\prime(s)}, \sigma_0^{\prime\prime(s)}), ~ j=1,\dots,D ~,~s=1,\dots,S$ between the occurring states and the average state, which gives the discussed base noise for detection.
	\item Draw a change sparsity $K$ uniformly as an integer number between $2$ and $D$. This gives the number of change affected dimensions.
	\item Determine which dimensions are affected by randomly drawing $K$ integer numbers uniformly between $1$ and $D$.
	\item Randomly draw the normal state $s^{*}$ the system is in after the change at time $t=\tau$ from the available states $S$ states.
	\item Draw a multiplicative change in correlation $a$ uniformly between $0$ and $1$. Then multiply the correlations between all variables $i$ and $d$ from the affected dimensions $\mathcal{D}$ with this change $a$ for $i\neq d$.
	\item Calculate $H_j(\mu_0^{\prime(s^{*})}, \sigma_0^{\prime(s^{*})},\mu_1^{\prime}, \sigma_1^{\prime}), ~ j=1,\dots,D$ between changed state and state, which gives the sensitivity for change detection with knowledge about the non-stationarity.
	\item Calculate $H_j(\overline{\mu}_0^{\prime\prime}, \overline{\sigma}_0^{\prime\prime},\mu_1^{\prime\prime}, \sigma_1^{\prime\prime}), ~ j=1,\dots,D$ between changed state and average state.
	\item Calculate the corrected Hellinger distance. This is done by subtracting the occurring additional noise (the maximum over all the distances between the states and the average state) from the Hellinger distance between changed state and average state: \\ $\mathcal{H}_j(\overline{\mu}_0^{\prime\prime}, \overline{\sigma}_0^{\prime\prime},\mu_1^{\prime\prime}, \sigma_1^{\prime\prime}) = \max( H_j(\overline{\mu}_0^{\prime\prime}, \overline{\sigma}_0^{\prime\prime},\mu_1^{\prime\prime}, \sigma_1^{\prime\prime}) - \max\limits_{s}(H_j(\overline{\mu}_0^{\prime\prime}, \overline{\sigma}_0^{\prime\prime},\mu_0^{\prime\prime(s)}, \sigma_0^{\prime\prime(s)})),0), ~ j=1,\dots,D$. \\
	This gives the sensitivity for change detection without knowledge about the non-stationarity.
	\item Repeat steps 4 to 10 for $10^3$ times.
	\item Repeat steps 1 to 11 for $10^3$ times. 
\end{enumerate}

\begin{figure}
		\centering
		\includegraphics[width=0.7\textwidth]{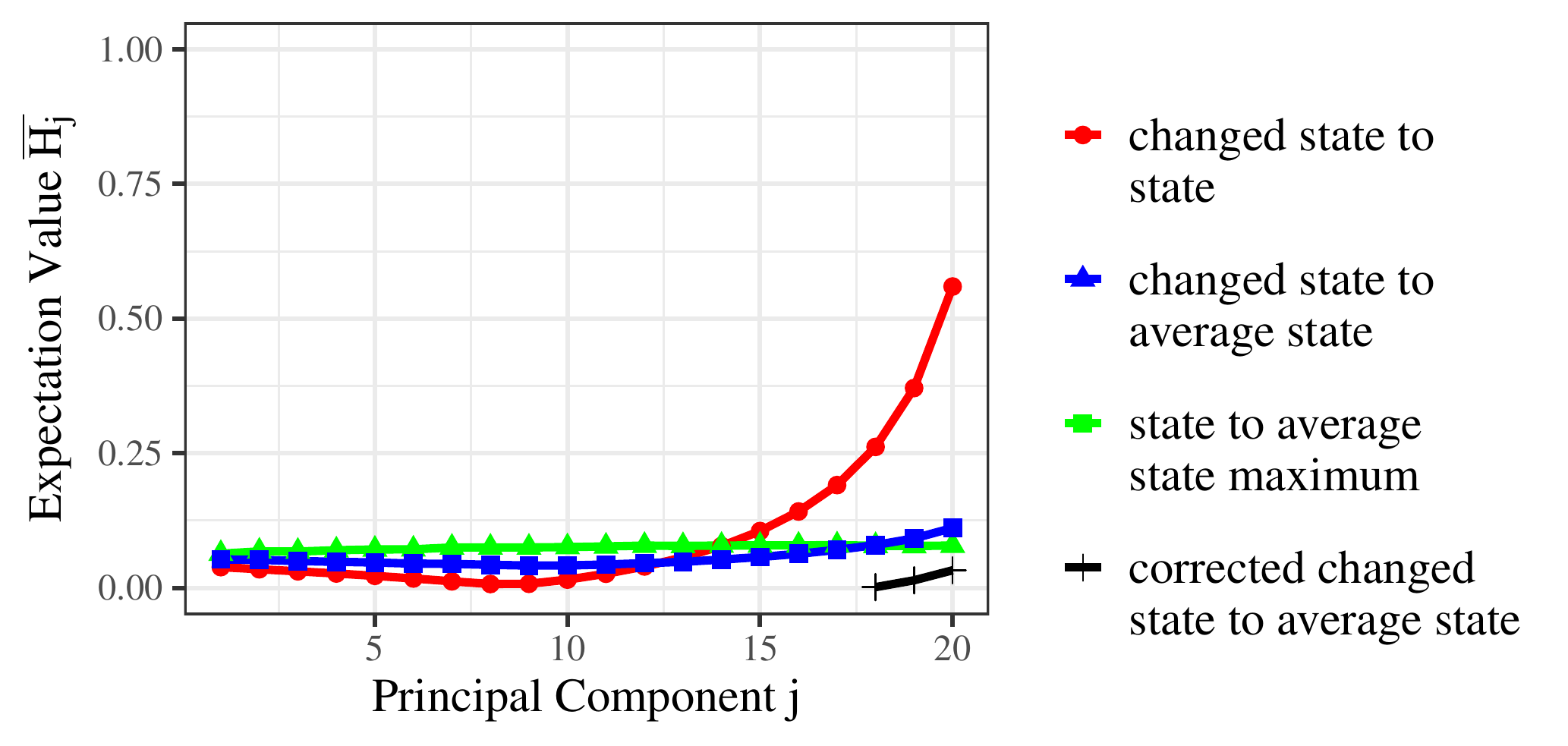}
		\caption{Monte Carlo estimates for the Hellinger distances of projections onto the different eigenvectors in presence of a change in correlation structure for $D=20$ and $S=3$. The original system states are created randomly and unrelated to each other. In color code the results for change sensitivity with knowledge of the non-stationarity (red), the uncorrected change sensitivity without knowledge (blue), the additional base noise induced by the system being in a state but comparing to the average state (green) and the corrected change sensitivity without knowledge (black) are shown. They correspond to the calculation steps 8, 9, 3 and 10 in the simulation, respectively.}
		\label{subfig:corresult_a}
	\end{figure}

In step 7. we only apply decreases in correlation. We do this to avoid many indefinite changed matrices \cite{Tveten2019}. One can easily imagine that a multiplicative increase could often lead to correlation coefficients larger than one. If any indefinite matrices still occur, we use Higham's algorithm \cite{Higham2002} to find the closest positive-definite one. The results of such a simulation for $D=20$ and $S=3$ are seen in Fig. \ref{subfig:corresult_a}. As mentioned before, correction of the Hellinger distance for the change detection with the average state is done inside the Monte Carlo runs. To obtain the single green line representing the correction values, we average over the Monte Carlo simulations. 

\begin{figure}
		\centering
		\includegraphics[width=0.7\textwidth]{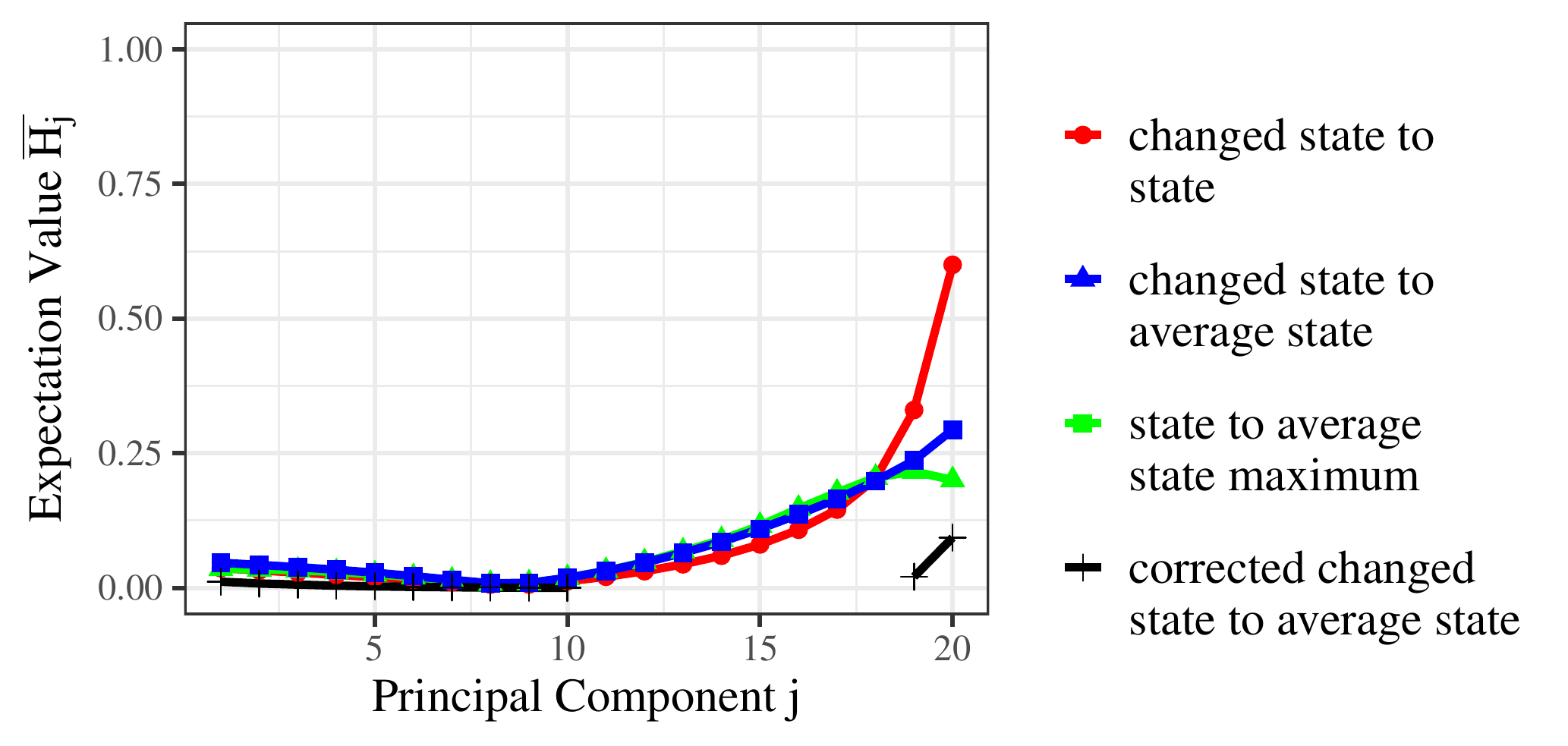}
		\caption{Monte Carlo estimates for the Hellinger distances of projections onto the different eigenvectors in presence of a change in correlation structure for $D=20$ and $S=3$. In the original system, one state is created randomly and the others are obtained from it by applying a change. In color code the results for change sensitivity with knowledge of the non-stationarity (red), the uncorrected change sensitivity without knowledge (blue), the additional base noise induced by the system being in a state but comparing to the average state (green) and the corrected change sensitivity without knowledge (black) are shown. They correspond to the calculation steps 8, 9, 3 and 10 in the simulation, respectively.}
		\label{subfig:corresult_b}
	\end{figure}

To compare states that emerged by applying changes to one random state, we simply change step 1. to the following substeps and present the results in Fig. \ref{subfig:corresult_b}:
\begin{enumerate}[label=\alph*)]
	\item Draw one random correlation matrix $C_0^{(1)}$ of dimension $D$ using the method described in \cite{Joe2006}.
	\item Draw a change sparsity $K$ uniformly as an integer number between $2$ and $D$. This gives the number of change affected dimensions.
	\item Determine which dimensions are affected by randomly drawing $K$ integer numbers uniformly between $1$ and $D$.
	\item Draw a multiplicative change in correlation $a$ uniformly between $0$ and $1$. Then multiply the correlations between all variables $i$ and $d$ from the affected dimensions $\mathcal{D}$ with this change $a$ for $i\neq d$.
	\item Repeat steps b) to d) $S-1$ times and use the changed correlation matrices as $C_0^{(s)}, ~ s=2,\dots,S$.
\end{enumerate}

 In both scenarios the sensitivity is greatest for the minor components in the changed state to state comparison. This is, of course, in accordance with the non-stationary results (see Fig. \ref{fig:oneState}) as we always compare to the correct normal state. The changed state to average state distance is larger than the one to the actual state for major components and a crossing point between the two appears towards minor components. This point lies with larger $j$ for the case of related original states. However, as pointed out before, we need to correct this Hellinger distance by the maximum distance between the actual states and the average state. This corrected Hellinger distance indicates the sensitivity of the change detection without knowledge about the non-stationarity. Its values are smaller than the ones with that knowledge for all principal components. In fact, they often lie below zero indicating no possible detection at all. The knowledge about non-stationarity greatly increases the possibility to detect changes.

\begin{figure}
		\centering
		\includegraphics[width=0.7\textwidth]{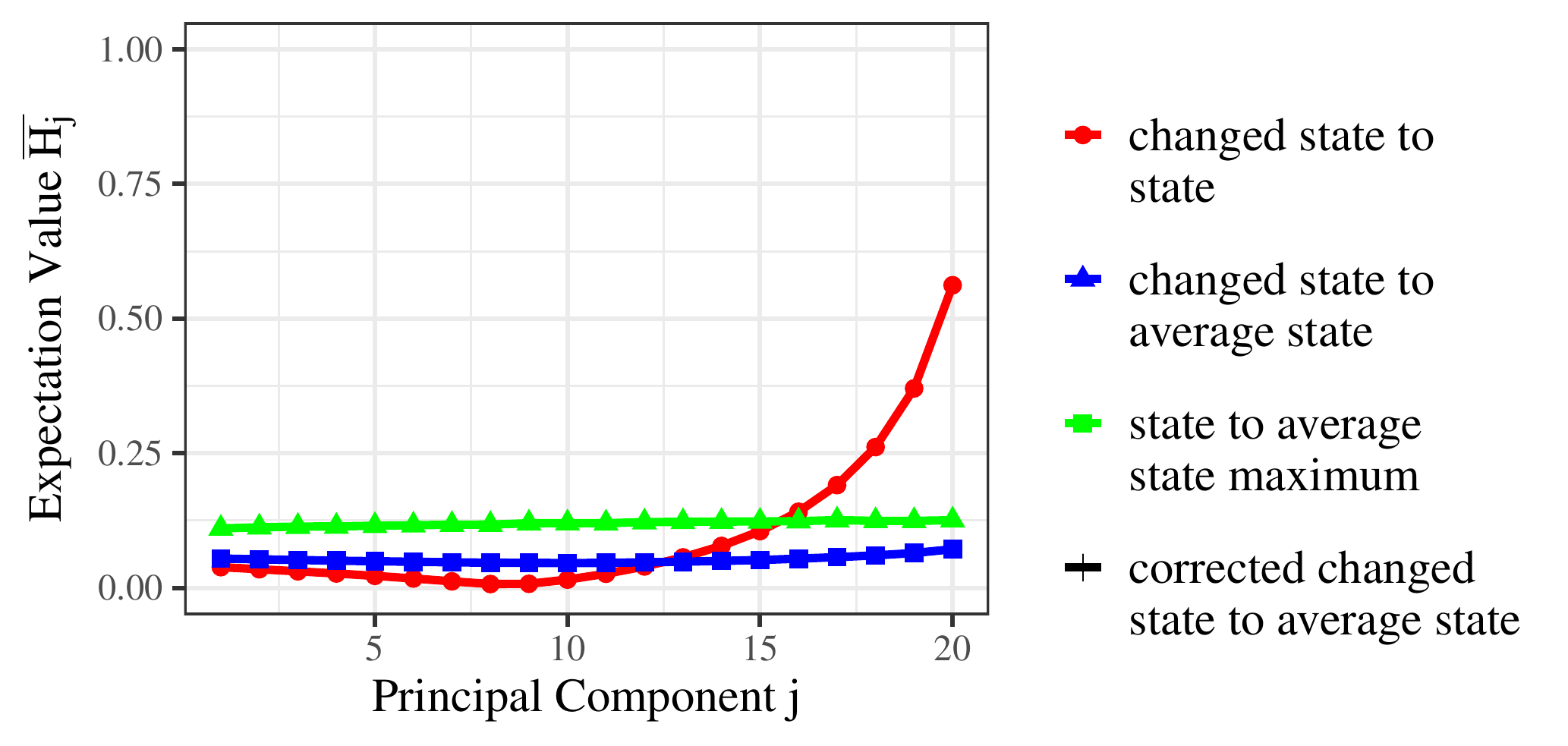}
		\caption{Monte Carlo estimates for the Hellinger distances of projections onto the different eigenvectors in presence of a change in correlation structure for $D=20$ and $S=7$. The original system states are created randomly and unrelated to each other. In color code the results for change sensitivity with knowledge of the non-stationarity (red), the uncorrected change sensitivity without knowledge (blue), the additional base noise induced by the system being in a state but comparing to the average state (green) and the corrected change sensitivity without knowledge (black) are shown. They correspond to the calculation steps 8, 9, 3 and 10 in the simulation, respectively.}
		\label{subfig:corresult_S7a}
	\end{figure}

\begin{figure}
		\centering
		\includegraphics[width=0.7\textwidth]{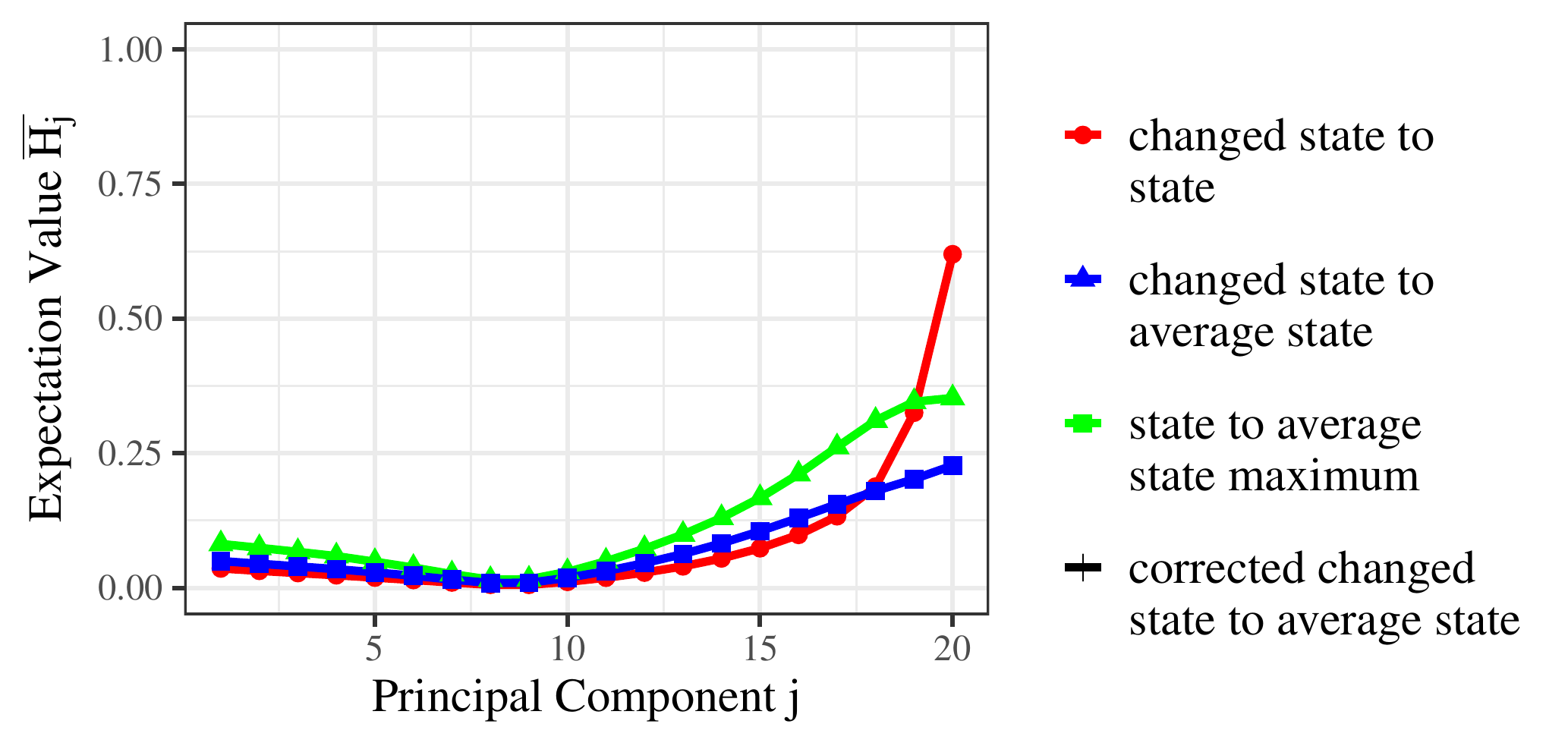}
		\caption{Monte Carlo estimates for the Hellinger distances of projections onto the different eigenvectors in presence of a change in correlation structure for $D=20$ and $S=7$. In the original system, one state is created randomly and the others are obtained from it by applying a change. In color code the results for change sensitivity with knowledge of the non-stationarity (red), the uncorrected change sensitivity without knowledge (blue), the additional base noise induced by the system being in a state but comparing to the average state (green) and the corrected change sensitivity without knowledge (black) are shown. They correspond to the calculation steps 8, 9, 3 and 10 in the simulation, respectively..}
		\label{subfig:corresult_S7b}
	\end{figure}

For the case of unrelated states the blue and green line seem to be almost flat, indicating that all components possess the same sensitivity. This is an inherent feature of the averaging. If the correlation matrices of the states are all entirely random, the correlation structures tend to cancel each other out. Simply put, the off-diagonal elements of the average matrix tend towards zero. This results in meaningless eigenvector structures. This is further underlined by Fig. \ref{subfig:corresult_S7a} showing the same results for $S=7$. With more states the average matrix is closer to zeros on the off-diagonal and the blue and green line are even flatter. It is one of the reasons for the introduction of the scenario with related states. Another, more application oriented, reason is that many systems will not change or reverse their entire behavior, but rather change the behavior of certain groups of variables. In this case, we see in Figs. \ref{subfig:corresult_S7a} and \ref{subfig:corresult_S7b} that with an increased number of states the detection without knowledge about the non-stationarity becomes impossible, i.e. the corrected Hellinger distances is always smaller than zero. For $S=3$ some detectability was given for minor principal components knowing just the average state, but this vanishes completely for $S=7$.

\begin{figure}
		\centering
		\includegraphics[width=0.7\textwidth]{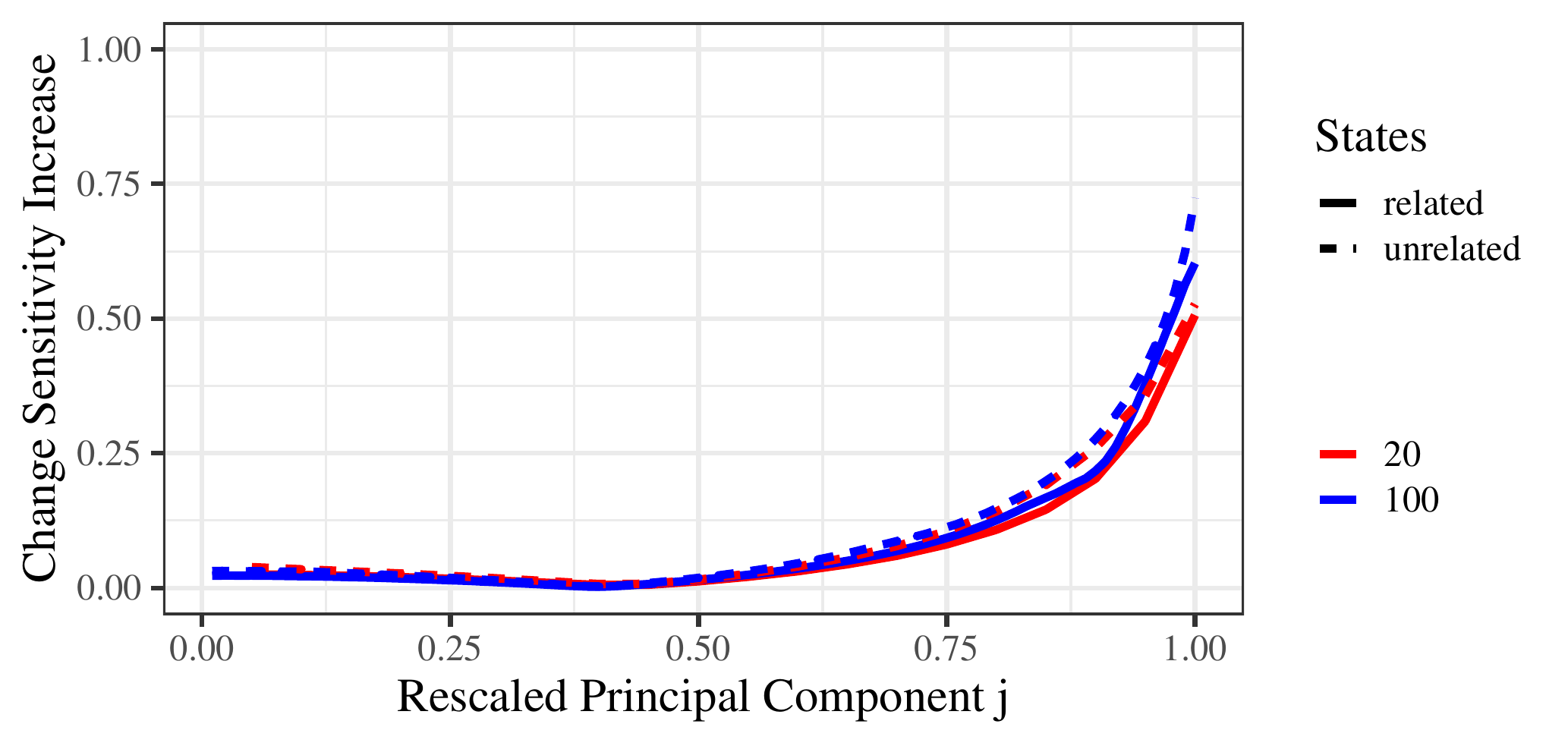}
		\caption{Sensitivity increase for a change in the correlation structure when knowledge about the state-wise non-stationarity is available compared to without that knowledge. Results are shown for different, color coded dimensions $D$, $S=3$ and unrelated random states (simulation scenario 1) as well as related states, that were generated from one random state (simulation scenario 2).}
		\label{subfig:corIncr_S3}
	\end{figure}

\begin{figure}
		\centering
		\includegraphics[width=0.7\textwidth]{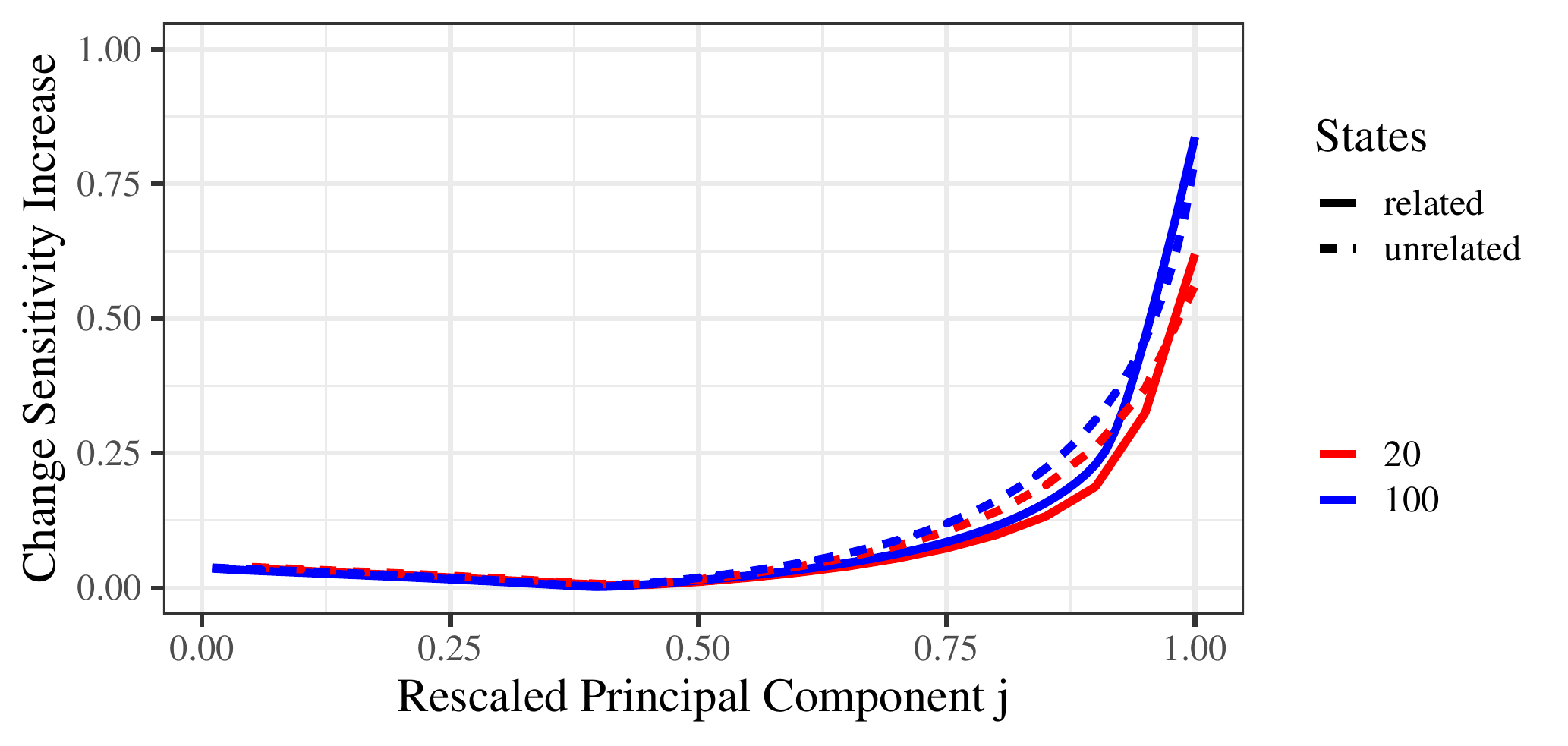}
		\caption{Sensitivity increase for a change in the correlation structure when knowledge about the state-wise non-stationarity is available compared to without that knowledge. Results are shown for different, color coded dimensions $D$, $S=7$ and unrelated random states (simulation scenario 1) as well as related states, that were generated from one random state (simulation scenario 2).}
		\label{subfig:corIncr_S7}
	\end{figure}

In general, the increase in change detectability with knowledge about non-stationarity compared to without is measured by the difference between the changed state to state distance and the corrected value for the average state. Results for this difference are shown in Figs. \ref{subfig:corIncr_S3} and \ref{subfig:corIncr_S7}. With non-stationarity present in the analysis, the minor components remain the most sensitive. Knowledge about the non-stationarity enables a more sensitive change detection for different numbers of normal states and different dimensions. As expected, the increase in sensitivity is larger for more states. It is also larger for unrelated normal states. Basically, these states are more different from each other than in the related case, so it is reasonable that the knowledge is more helpful here. We see, however, that for $S=7$ for the largest $j$, where detectability is highest in general, the increase is larger for related normal states. As seen in Figs. \ref{subfig:corresult_S7a} and \ref{subfig:corresult_S7b} for $S=7$ the corrected state to average state distance is always zero for both scenarios. So any difference in the sensitivity increase between the scenarios can only stem from a difference in the changed state to state distance. This difference exists purely due to a technicality: Because correlation coefficients cannot be larger than 1 the changes applied here are multiplicative between 0 and 1, i.e. they only reduce correlation. As this is also true for the changes performed to obtain the related states, the related normal states have weaker correlations on average than in the scenario with random states. This leads to a change of the sensitivity in the changed state to state scenario in Figs. \ref{subfig:corresult_b} and \ref{subfig:corresult_S7b}. Here, the results are different from the stationary case, because the underlying set of matrices is changed. In general however, as long as different normal states exist, the knowledge about non-stationarity increases change detection sensitivity. As this is the main interest of the current study, we did not pursue this effect further, but it is interesting for future studies. Moving on, the difference in $D$ does not change the overall results. For small $j$ the increase is a bit larger for smaller dimensions, whereas a for large $j$ it is the other way around. With the sensitivity for change being greatest for the minor components, the knowledge about non-stationarity is very important also in high-dimensional systems. It is noteworthy that an increase in the dimension of the matrix to $D=100$ means that the relative change sparsity can be much higher as the smallest value of changed dimensions remains at two independent of $D$. Our results are also valid for sparse changes.

In summary, the knowledge about state-wise non-stationarity is important for the detection of changes. The sensitivity increase with knowledge is largest for the minor components, where change detection without non-stationarity is already most sensitive. For the major components the increase is still detectable, but much smaller. We could speculate that PCA for dimension reduction in a system is therefore still quite possible without knowing about the non-stationarity, but this cannot be tested or verified in our setup and is not in the scope of this paper.



\subsection{Change in mean}
\label{sec:meanresults}

We now analyze the effect of changes in the mean values of variables. Again, we perform Monte Carlo simulations with various change scenarios to obtain results. The procedure in general is similar to the one used for changes in correlation, but the details change. We also simulate two different scenarios of normal states again: related and unrelated. In the case of unrelated, random states our simulation follows the steps:
\begin{enumerate}
	\item Draw a random correlation matrix $C_0$ of dimension $D$ using the method described in \cite{Joe2006}, which is the same for all states.
	\item Draw $S$ vectors of means $\bm{\mu}_0^{(s)}$, where each element is uniformly drawn as a non-integer value between $-3$ and $3$.
	\item Calculate the element-wise average vector of means $\bm{\overline{\mu}}_0$ before the change.
	\item Assume standardization with average pre-change parameters and calculate $H_j(\tilde{\overline{\mu}}_0^{\prime\prime}, \tilde{\overline{\sigma}}_0^{\prime\prime},\tilde{\mu}_0^{\prime\prime(s)}, \tilde{\sigma}_0^{\prime\prime(s)}), ~ j=1,\dots,D~,~s=1,\dots,S$ between the states and the average state, which gives the discussed base noise for detection.
	\item Draw a change sparsity $K$ uniformly as an integer number between $2$ and $D$. This gives the number of change affected dimensions.
	\item Determine which dimensions are affected by randomly drawing $K$ integer numbers uniformly between $1$ and $D$.
	\item Randomly draw the normal state $s^{*}$ the system is in after the change at time $t=\tau$ from the available states $S$ states.
	\item Draw an additive change in mean $\Delta \mu$ uniformly between $-3$ and $3$. To obtain $\bm{\mu}_1$, i.e. the vector of means after the change, from the vector of means of the original state $\bm{\mu}_{0}^{(s^{*})}$ the change value $\Delta \mu$ is added to the elements of the affected dimensions.
	\item Assume standardization with known state pre-change parameters, i.e. $\bm{\hat{\mu}}_{0}^{(s^{*})} = \bm{0}$ and
	$$\left(\bm{\hat{\mu}}_1\right)_d =
	\begin{cases}
		\Delta \mu,& ~ \text{if } d \text{ is an affected dimension} \\
		0, & ~ \text{otherwise}
	\end{cases},$$
	and calculate  $H_j(\hat{\mu}_0^{\prime(s^{*})}, \hat{\sigma}_0^{\prime(s^{*})},\hat{\mu}_1^{\prime}, \hat{\sigma}_1^{\prime}), ~ j=1,\dots,D$ between changed state and state. This gives the sensitivity for change detection with knowledge about the non-stationarity.
	\item Assume standardization with average pre-change parameters, i.e. $\bm{\tilde{\overline{\mu}}}_0=\bm{0}$ and
	$$\left(\bm{\tilde{\mu}}_1\right)_d =
	\begin{cases}
		(\bm{\mu}_{0}^{(s^{*})})_d - (\bm{\overline{\mu}}_{0})_d + \Delta \mu,& ~ \text{if } d \text{ is an affected dimension} \\
		(\bm{\mu}_{0}^{(s^{*})})_d - (\bm{\overline{\mu}}_{0})_d, & ~ \text{otherwise}
	\end{cases},$$
	and calculate $H_j(\tilde{\overline{\mu}}_0^{\prime\prime}, \tilde{\overline{\sigma}}_0^{\prime\prime},\tilde{\mu}_1^{\prime\prime}, \tilde{\sigma}_1^{\prime\prime}), ~ j=1,\dots,D$ between changed state and average state.
	\item Calculate the corrected Hellinger distance. This is done by subtracting the occurring additional noise (the maximum over all the distances between the states and the average state) from the Hellinger distance between changed state and average state: \\
	$\mathcal{H}_j(\tilde{\overline{\mu}}_0^{\prime\prime}, \tilde{\overline{\sigma}}_0^{\prime\prime},\tilde{\mu}_1^{\prime\prime}, \tilde{\sigma}_1^{\prime\prime}) = \max( H_j(\tilde{\overline{\mu}}_0^{\prime\prime}, \tilde{\overline{\sigma}}_0^{\prime\prime},\tilde{\mu}_1^{\prime\prime}, \tilde{\sigma}_1^{\prime\prime}) - \max\limits_{s}(H_j(\tilde{\overline{\mu}}_0^{\prime\prime}, \tilde{\overline{\sigma}}_0^{\prime\prime},\tilde{\mu}_0^{\prime\prime(s)}, \tilde{\sigma}_0^{\prime\prime(s)})),0), ~ j=1,\dots,D$.
	\item Repeat steps 5 to 11 for $10^3$ times.
	\item Repeat steps 1 to 12 for $10^3$ times. 
\end{enumerate}

To have states that emerged by applying changes to one random state, we simply change step 2. to the substeps:
\begin{enumerate}[label=\alph*)]
	\item Draw one vector of means $\bm{\mu}_0^{(1)}$, where each element is uniformly drawn as a non-integer value between $-3$ and $3$.
	\item Draw a change sparsity $K$ uniformly as an integer number between $2$ and $D$. This gives the number of change affected dimensions.
	\item Determine which dimensions are affected by randomly drawing $K$ integer numbers uniformly between $1$ and $D$.
	\item Draw an additive change in mean $\Delta \mu$ uniformly between $-3$ and $3$. To obtain $\bm{\mu}_0^{(s)}, ~ s \neq 1$, i.e. the vector of means of a state $s$, from the vector of means of state $1$ $\bm{\mu}_{0}^{(1)}$ the change value $\Delta \mu$ is added to the elements of the affected dimensions.
	\item Repeat steps b) to d) $S-1$ times and use the changed vectors of means as $\bm{\mu}_0^{(s)}, ~ s=2,\dots,S$.
\end{enumerate}

\begin{figure}
		\centering
		\includegraphics[width=0.7\textwidth]{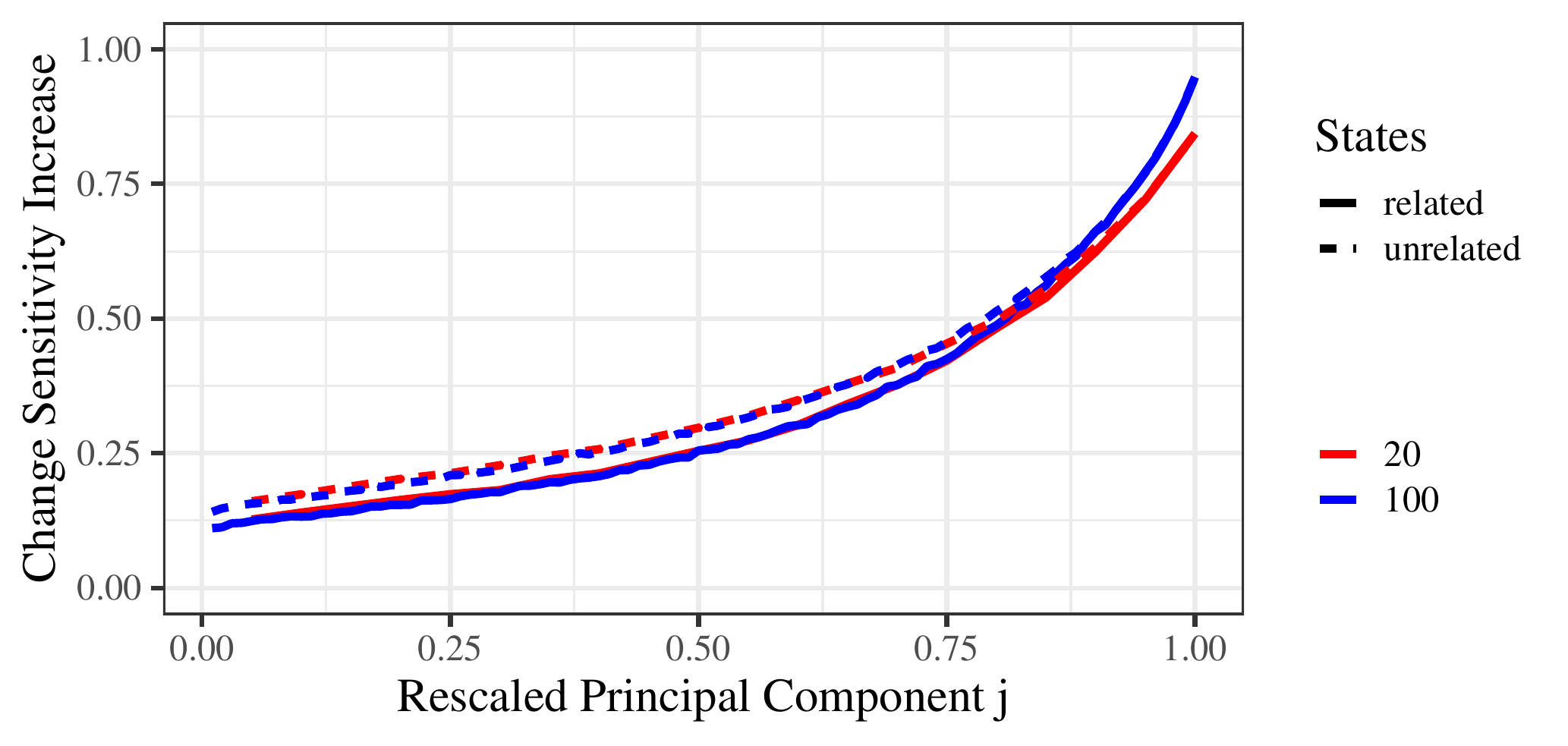}
		\caption{Sensitivity increase for a change in mean when knowledge about the state-wise non-stationarity is available compared to without that knowledge. Results are shown for different, color coded dimensions $D$, $S=3$ and unrelated random states (simulation scenario 1) as well as related states, that were generated from one random state (simulation scenario 2).}
		\label{subfig:meanIncr_S3}
	\end{figure}

\begin{figure}
		\centering
		\includegraphics[width=0.7\textwidth]{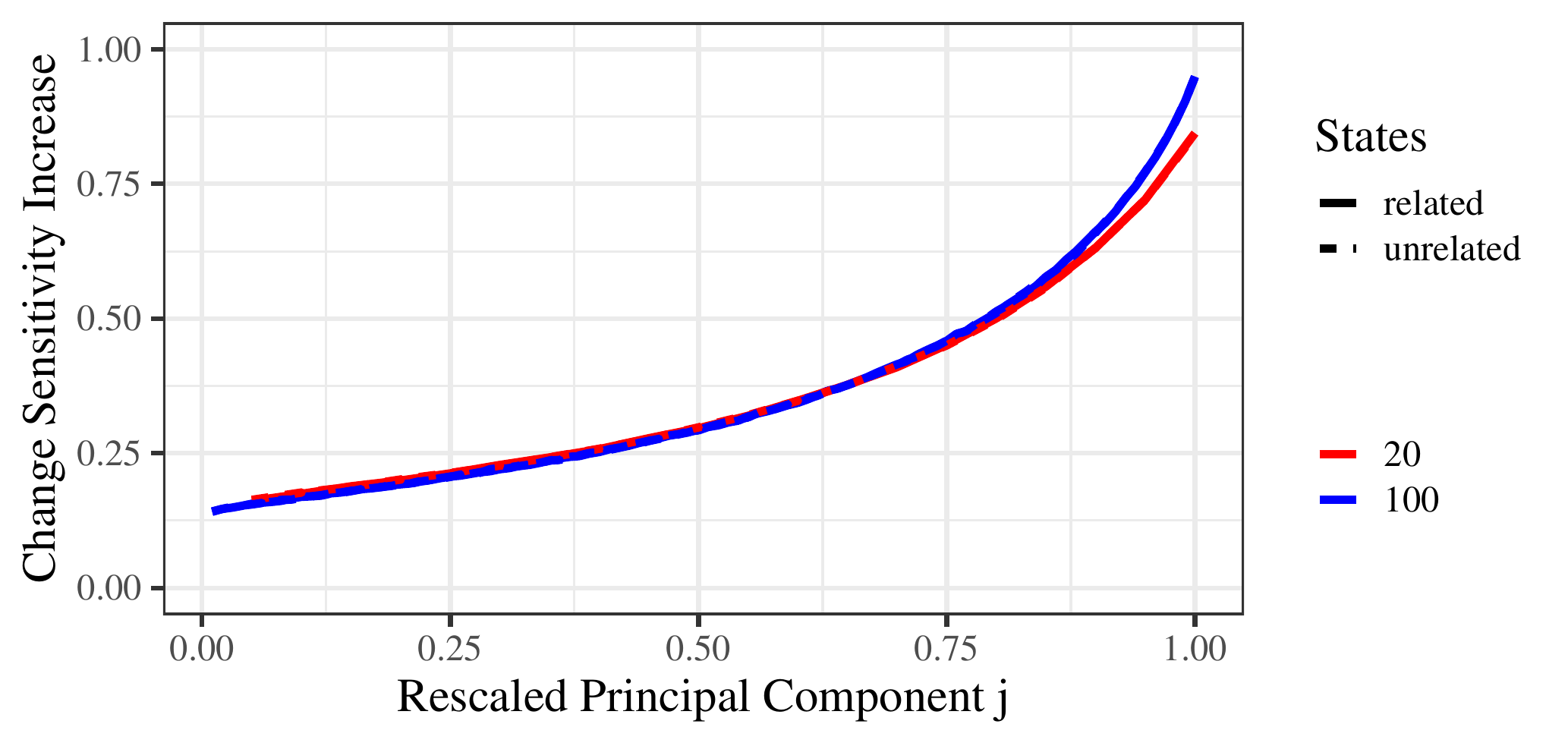}
		\caption{Sensitivity increase for a change in mean when knowledge about the state-wise non-stationarity is available compared to without that knowledge. Results are shown for different, color coded dimensions $D$, $S=7$ and unrelated random states (simulation scenario 1) as well as related states, that were generated from one random state (simulation scenario 2).}
		\label{subfig:meanIncr_S7}
	\end{figure}

As we saw in section \ref{sec:corresults} the most interesting result is the increase in sensitivity with knowledge about the non-stationarity compared to change detection without said knowledge. These simulation results are shown in Figs. \ref{subfig:meanIncr_S3} and \ref{subfig:meanIncr_S7}. We see a clear increase in sensitivity for all simulated parameters. As before with changes in correlation structure, the knowledge about non-stationarity is slightly more important when the original normal states are related to each other and therefore not entirely different for $S=3$. For $S=7$ this difference is no longer visible. This is because the corrected Hellinger distances are always zero here and the changed state to state distances are not substantially different between the two scenarios. There is no influential change in the underlying set of states in contrast to the one seen in Sec. \ref{sec:corresults}.

We again conclude that the non-stationarity is in general important for change detection. The increase seems to be more important for small $j$, i.e. major components, as compared to the results in section \ref{sec:corresults}. However, this is most likely due to the fact that the change sensitivity without non-stationarity already exhibits the same changes in their dependency on $j$ (see Fig. \ref{fig:oneState}).


\subsection{Change in standard deviation}
\label{sec:sdresults}

Finally, we analyze the sensitivity for changes in standard deviation in the presence of non-stationarity. The basic Monte Carlo process is the same, but once more the details are different. We simulate related and unrelated normal states again. In the case of unrelated, random states our simulation follows the steps:
\begin{enumerate}
	\item Draw a random correlation matrix $C_0$ of dimension $D$ using the method described in \cite{Joe2006}, which is the same for all states.
	\item Draw $S$ vectors of standard deviations $\bm{\sigma}_0^{(s)}$, where each element is uniformly drawn as a non-integer value between $1/3$ and $2$.
	\item Calculate the element-wise average vector of variances $\bm{\overline{\sigma}}_0^2$ before the change\footnote{We discussed in section \ref{sec:theory} that simply taking the average is probably closest to a real use case. In the case of non-changing means between the states and with every state appearing for the same amount time, this actually gives the correct variance over all states.}. Then take the square root to get the average standard deviations.
	\item Assume standardization with average pre-change parameters. Then $\tilde{C}_0$ is a well-defined correlation matrix with ones on its diagonal. To assume normalization with average parameters in the state and calculate $\tilde{C}_{0}^{(s)}$, we need the state covariance matrix. It is obtained by undoing the correct normalization of the state correlation matrix $$\Sigma_0^{(s)} = \mathrm{diag}\left(\bm{\sigma}_{0}^{(s)}\right) \hat{C}_0^{(s)} \mathrm{diag}\left(\bm{\sigma}_{0}^{(s)}\right).$$
	It is then wrongly normalized with the average pre-change parameters $$\tilde{C}_0^{(s)} = \left(\mathrm{diag}\left(\bm{\overline{\sigma}}_{0}\right)\right)^{-1} \Sigma_0^{(s)} \left(\mathrm{diag}\left(\bm{\overline{\sigma}}_{0}\right)\right)^{-1}.$$ $\tilde{C}_{0}^{(s)}$ is not a well-defined correlation matrix. We then calculate the Hellinger distance $H_j(\tilde{\overline{\mu}}_0^{\prime\prime}, \tilde{\overline{\sigma}}_0^{\prime\prime},\tilde{\mu}_0^{\prime\prime(s)}, \tilde{\sigma}_0^{\prime\prime(s)}), ~ j=1,\dots,D~,~s=1,\dots,S$ between the states and the average state, which gives the discussed base noise for detection.
	\item Draw a change sparsity $K$ uniformly as an integer number between $2$ and $D$. This gives the number of change affected dimensions.
	\item Determine which dimensions are affected by randomly drawing $K$ integer numbers uniformly between $1$ and $D$.
	\item Randomly draw the normal state $s^{*}$ the system is in after the change at time $t=\tau$ from the available states $S$ states.
	\item Randomly decide if the change increases or decreases the standard deviation. Then, draw a multiplicative change in standard deviation $\Delta \sigma$ uniformly between $1$ and $3$ or between $1/3$ and $1$, respectively. $\bm{\sigma}_1$ is obtained by multiplying the elements of $\bm{\sigma}_{0}^{(s^{*})}$ with $\Delta \sigma$ for the affected dimensions.
	\item Assume standardization with known state pre-change parameters. Then $\hat{C}_0^{(s^{*})}$ is a well-defined correlation matrix with ones on its diagonal and $\hat{C}_1$ is calculated analogous to the description in step 4. Then calculate the Hellinger distance  $H_j(\hat{\mu}_0^{\prime(s^{*})}, \hat{\sigma}_0^{\prime(s^{*})},\hat{\mu}_1^{\prime}, \hat{\sigma}_1^{\prime}), ~ j=1,\dots,D$ between changed state and state. This gives the sensitivity for change detection with knowledge about the non-stationarity.
	\item Assume standardization with average pre-change parameters. Then $\tilde{\overline{C}}_0$ is a well defined correlation matrix and $\tilde{C}_1$ can be calculated analogous to the procedure in step 4. Then calculate the Hellinger distance $H_j(\tilde{\overline{\mu}}_0^{\prime\prime}, \tilde{\overline{\sigma}}_0^{\prime\prime},\tilde{\mu}_1^{\prime\prime}, \tilde{\sigma}_1^{\prime\prime}), ~ j=1,\dots,D$ between changed state and average state.
	\item Calculate the corrected Hellinger distance. This is done by subtracting the occurring additional noise (the maximum over all the distances between the states and the average state) from the Hellinger distance between changed state and average state:  \\
	$\mathcal{H}_j(\tilde{\overline{\mu}}_0^{\prime\prime}, \tilde{\overline{\sigma}}_0^{\prime\prime},\tilde{\mu}_1^{\prime\prime}, \tilde{\sigma}_1^{\prime\prime}) = \max(H_j(\tilde{\overline{\mu}}_0^{\prime\prime}, \tilde{\overline{\sigma}}_0^{\prime\prime},\tilde{\mu}_1^{\prime\prime}, \tilde{\sigma}_1^{\prime\prime}) - \max\limits_{s}(H_j(\tilde{\overline{\mu}}_0^{\prime\prime}, \tilde{\overline{\sigma}}_0^{\prime\prime},\tilde{\mu}_0^{\prime\prime(s^{*})}, \tilde{\sigma}_0^{\prime\prime(s^{*})})),0), ~ j=1,\dots,D$.
	\item Repeat steps 5 to 11 for $10^3$ times.
	\item Repeat steps 1 to 12 for $10^3$ times.
\end{enumerate}

To have states that emerged by applying changes to one random state, we simply change step 2. to the substeps:
\begin{enumerate}[label=\alph*)]
	\item Draw one vector of standard deviations $\bm{\sigma}_0^{(1)}$, where each element is uniformly drawn as a non-integer value between $1/3$ and $2$.
	\item Draw a change sparsity $K$ uniformly as an integer number between $2$ and $D$. This gives the number of change affected dimensions.
	\item Determine which dimensions are affected by randomly drawing $K$ integer numbers uniformly between $1$ and $D$.
	\item Randomly decide if new state has increased or decreased standard deviation. Then, draw a multiplicative change in standard deviation $\Delta \sigma$ uniformly between $1$ and $3$ or between $1/3$ and $1$, respectively. $\bm{\sigma}_0^{(s)}, ~s\neq1$ is obtained by multiplying the elements of $\bm{\sigma}_{0}^{(1)}$ with $\Delta \sigma$ for the affected dimensions.
	\item Repeat steps b) to d) $S-1$ times and use the changed vectors of standard deviations as $\bm{\sigma}_0^{(s)}, ~ s=2,\dots,S$.
\end{enumerate}

\begin{figure}
		\centering
		\includegraphics[width=0.7\textwidth]{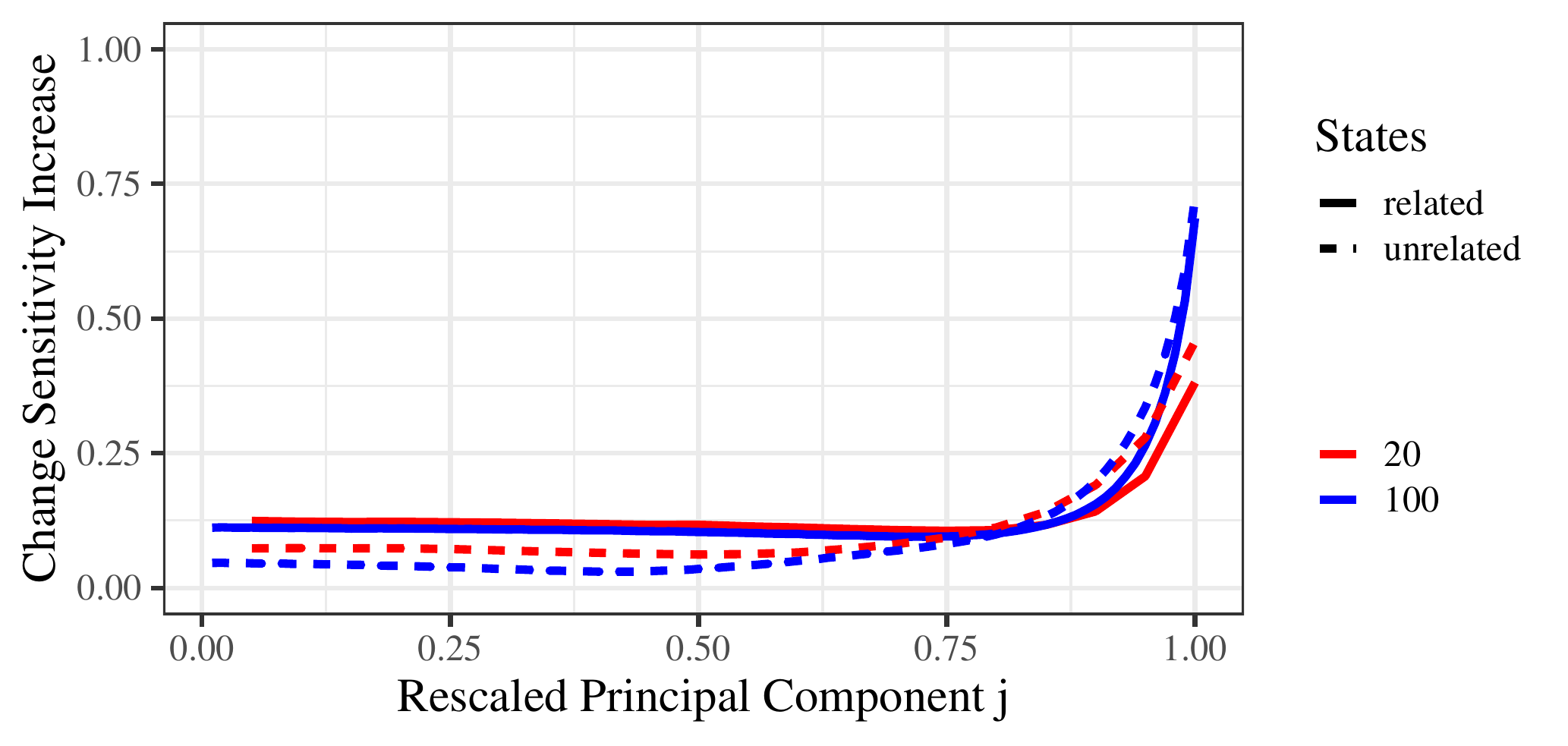}
		\caption{Sensitivity increase for a change in standard deviation when knowledge about the state-wise non-stationarity is available compared to without that knowledge. Results are shown for different, color coded dimensions $D$, $S=3$ and unrelated random states (simulation scenario 1) as well as related states, that were generated from one random state (simulation scenario 2).}
		\label{subfig:varIncr_S3}
	\end{figure}

\begin{figure}
		\centering
		\includegraphics[width=0.7\textwidth]{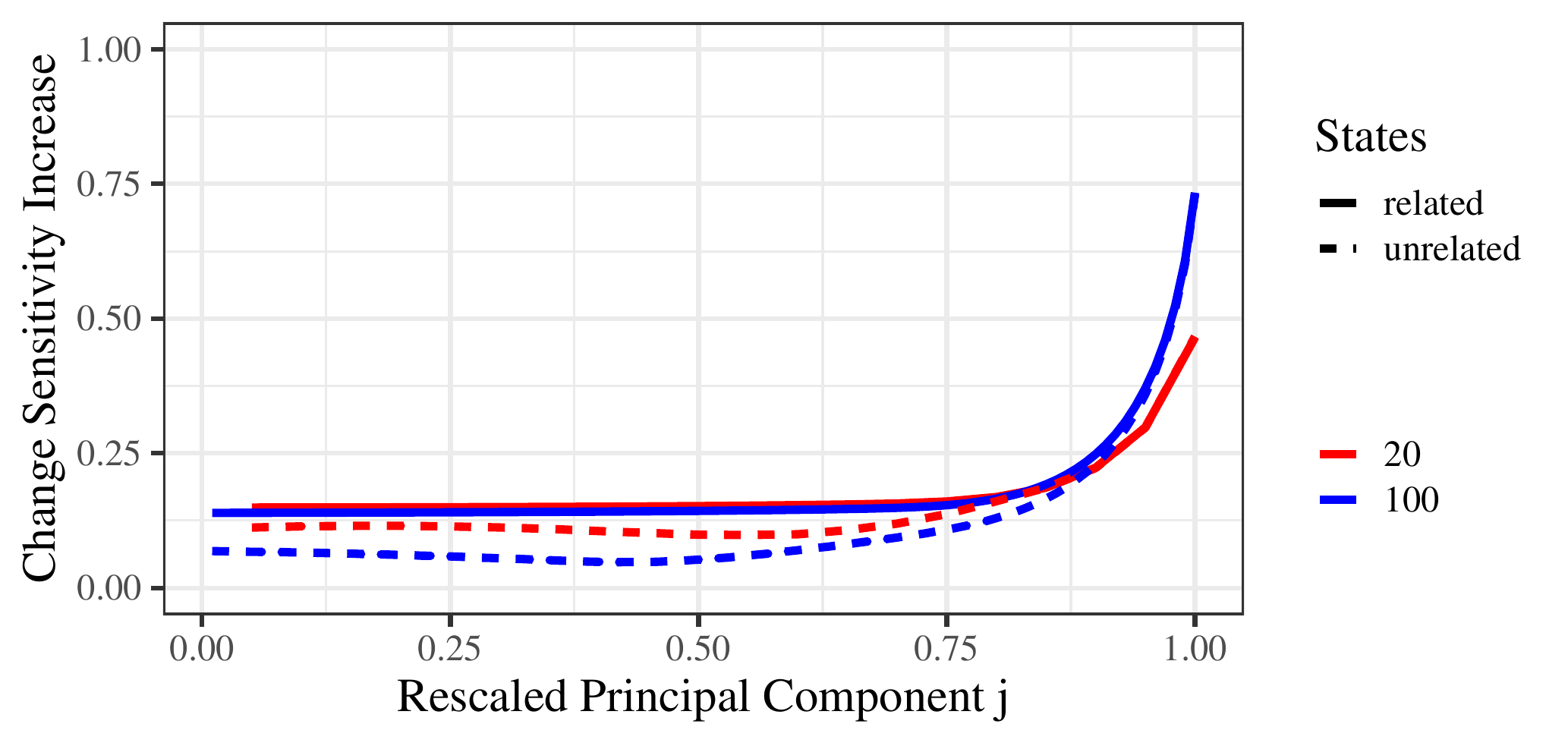}
		\caption{Sensitivity increase for a change in standard deviation when knowledge about the state-wise non-stationarity is available compared to without that knowledge. Results are shown for different, color coded dimensions $D$, $S=7$ and unrelated random states (simulation scenario 1) as well as related states, that were generated from one random state (simulation scenario 2).}
		\label{subfig:varIncr_S7}
	\end{figure}

The results for detection sensitivity increase due to knowledge about the non-stationarity are shown in Figs. \ref{subfig:varIncr_S3} and \ref{subfig:varIncr_S7}. It seems to be small for a large number of components $j$, especially the major components. As was the case with a comparison between changes in mean and correlation structure, this is largely due to the different sensitivity in the stationary case (see Fig. \ref{fig:oneState}). The curve of the increase always shows a similar behavior over $j$ as the stationary sensitivity itself shows. This underlines the importance of knowledge about non-stationarity as one would use these components for practical applications. Changes in standard deviation are almost impossible to detect for small $j$ even in the stationary case. Therefore, our results cannot show a large increase in that regime. For large $j$ we can again conclude that the knowledge about the non-stationarity of the system is quite important and becomes even more so if the original states are similar to each other. Again, as with results for changes in mean, the difference between related and unrelated states is smaller for $S=7$ and even vanished for large $j$ in that case. This is again due to a vanishing corrected Hellinger distance between state and average stat in this regime. 


\section{Application to Real Data: Traffic as an Example}
\label{sec:trafficexample}

To demonstrate how the method works, we give an instructive example with real data collected on German motorways. We look at 35 cross-sections (detectors) on the Cologne orbital motorway in 2015. The measured values are flows of vehicles, i.e. the number of vehicles passing a detector per time, in the counterclockwise direction. The time resolution of the data is one minute. As the orbital motorway consists of sections with different numbers of lanes, we do not look at a single lane, but rather at the flow accumulated over all lanes. The flow at time $t$ at cross-section $k$ is denoted $q_k (t)$. The dataset is described in more detail in \cite{WangShanshan2020}.

This example is chosen because it allows a simple definition of "normal" and "non-normal", i.e. changes in the system. These terms serve only as labels without further interpretation. We consider as normal the workdays Monday through Thursday as they present the typical rush hour behavior one expects from traffic. Fridays are excluded as the afternoon rush hour is stretched in time due to a strong variation in the ends of the working hours. Furthermore, we exclude the North Rhine-Westphalia bank holidays. Accordingly, we can afterwards try to detect weekend or holiday behavior as non-normal. Another advantage of this example is that everybody is aware of the non-stationarity of the system. Traffic is very different depending on the time of day. This means the non-stationarity in this system is not governed by a fluctuating variable but simply by the time of day.

\begin{figure}
	\centering
	\begin{subfigure}{0.45\textwidth}
		\includegraphics[width=\textwidth]{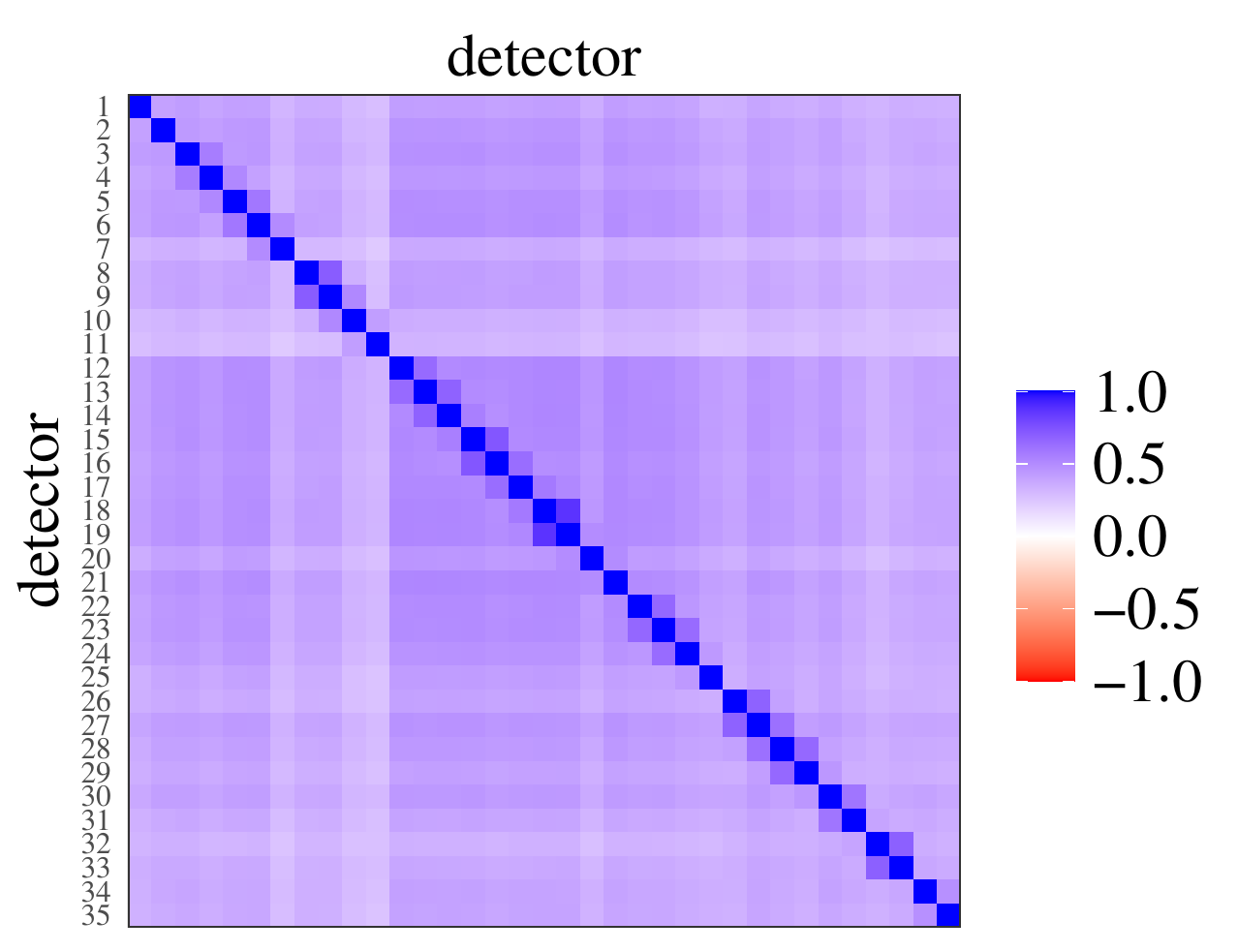}
		\caption{Cluster 1}
		\label{subfig:clustercenter1}
	\end{subfigure}
	\begin{subfigure}{0.45\textwidth}
		\includegraphics[width=\textwidth]{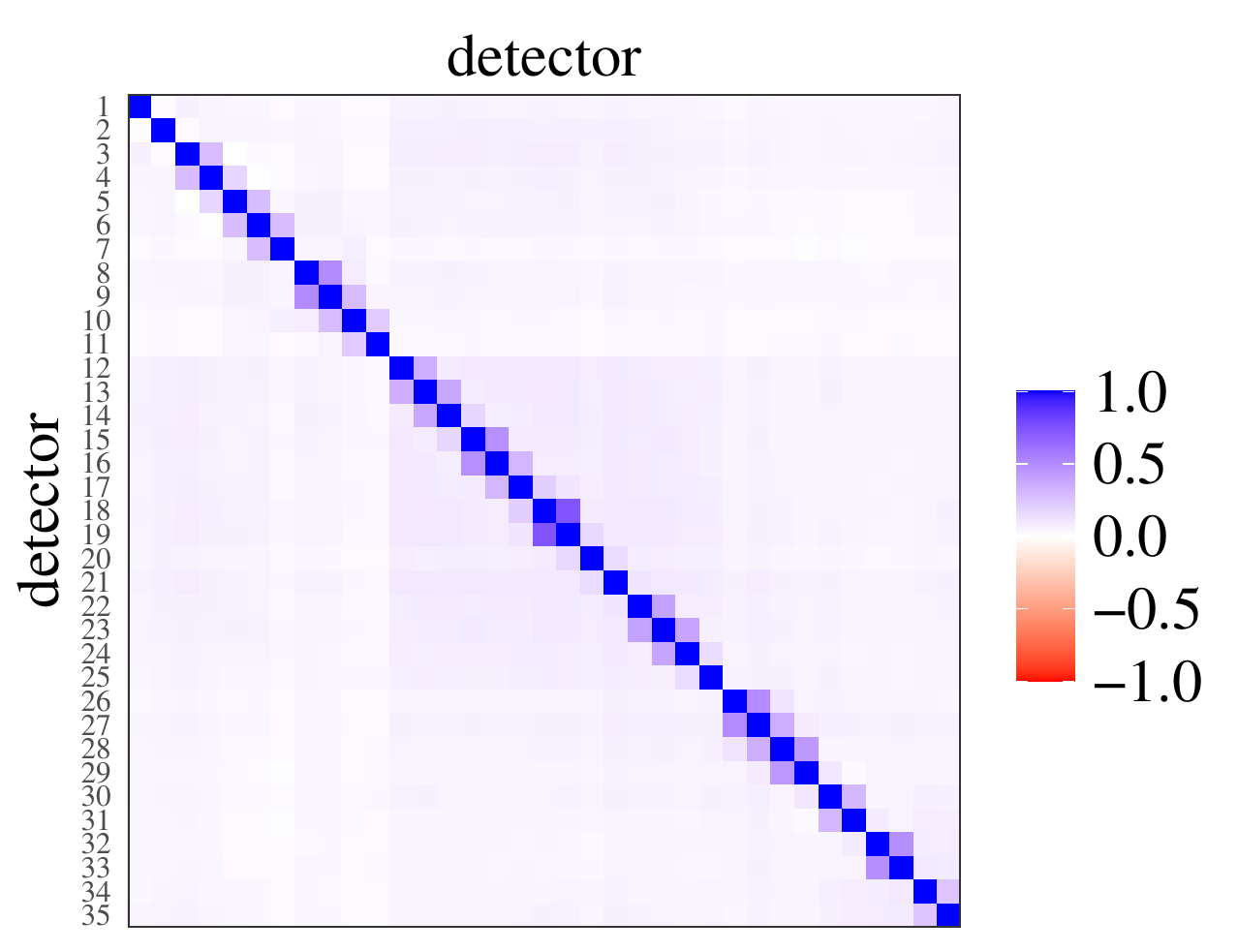}
		\caption{Cluster 2}
		\label{subfig:clustercenter2}
	\end{subfigure}
	\caption{Correlation matrix cluster centers calculated as element-wise means over all matrices sorted into a cluster. X-axis and y-axis both show the different traffic detectors, but labels were removed on the x-axis for better readability of the figure. Each matrix element is the mean Pearson correlation coefficient between the traffic flow signals of two detectors and its value is color coded.}
	\label{fig:clustercenters}
\end{figure}

\begin{figure}
	\centering
	\includegraphics[width=0.7\textwidth]{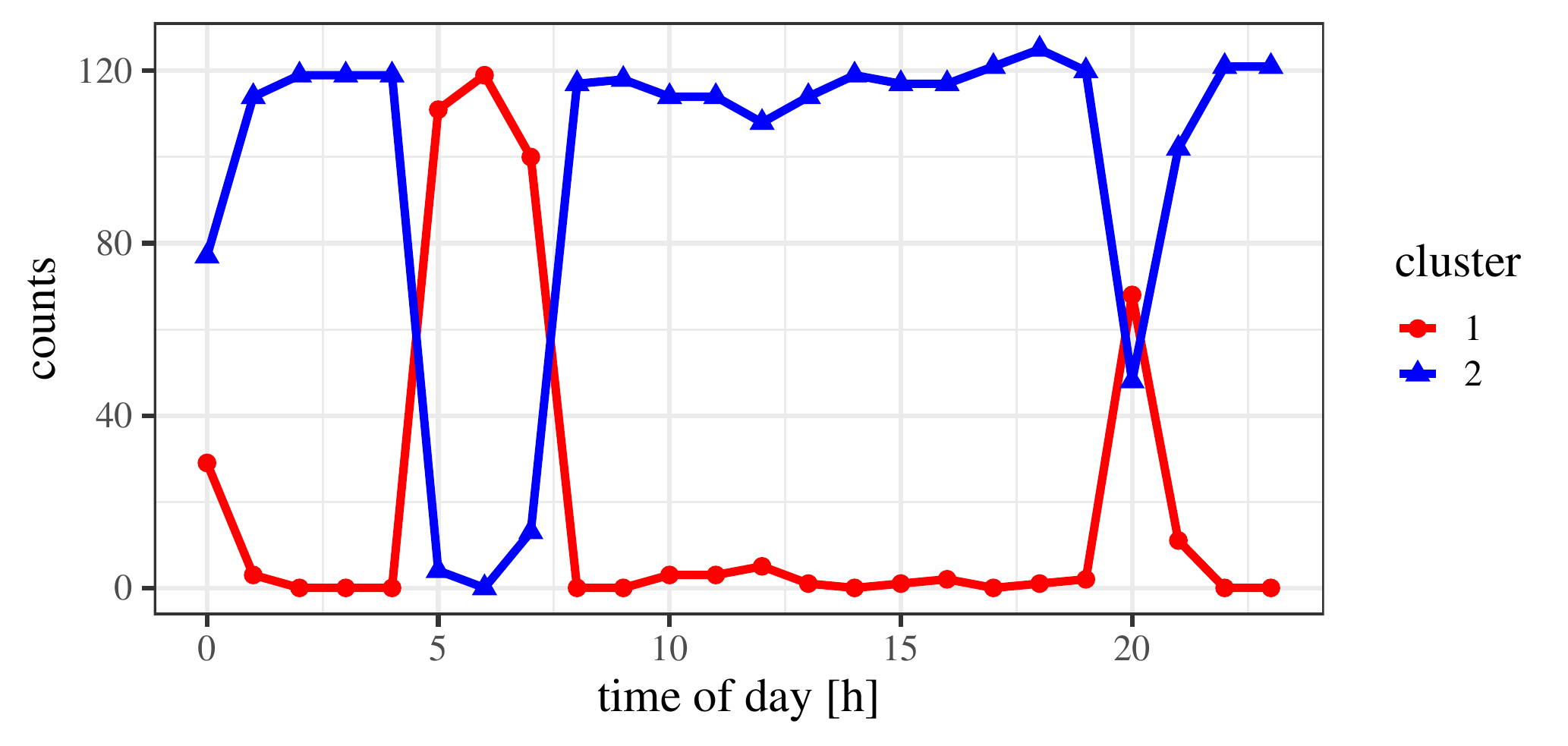}
	\caption{Histogram counts of the appearances of cluster 1 and 2 during the 24 hours of a day. The histogram is calculated over all used week days with a bin width of 1 hour.}
	\label{fig:clusterOverTime}
\end{figure}

The first thing one has to do, to test the proposed method on real data is classify the non-stationarity in the data. There is, of course, not only one way to do this. Here, we calculate correlation matrices for each hour of the normal workdays. This is done with non-overlapping time windows called epochs with a length $\tau_{\mathrm{ep}} = 1\si{h}$. A day will thus contain 24 correlation matrices. The matrix for time $t=\tau$ is calculated from the data $q_k(t), ~ k=1,\dots,35 ~, ~ \tau-\tau_{\mathrm{ep}} < t \leq \tau$. We write $| \tau_{\mathrm{ep}} |$ to denote the number of data points during this epoch. We normalize for each detector $k$ and each time window separately to zero mean and standard deviation one:
\begin{equation}
	Q_k(t) = \frac{q_k(t)-\langle q_k(t)\rangle_{\tau}}{\sqrt{\langle q_k^2(t) \rangle_{\tau} - \langle q_k(t)\rangle^2_{\tau}}}, ~ \tau-\tau_{\mathrm{ep}} < t \leq \tau.
\end{equation}
Here, $\langle \dots \rangle_{\tau}$ denotes the average in the one hour epoch. We arrange the data in a $35 \times |\tau_{\mathrm{ep}}|$ matrix
\begin{equation}
	Q(\tau) = \begin{bmatrix}
		Q_1(\tau- \tau_{\mathrm{ep}}) & \dots & Q_1(\tau) \\
		\vdots & & \vdots \\
		Q_k(\tau- \tau_{\mathrm{ep}}) & \ddots & Q_k^{(l)}(\tau) \\
		\vdots & & \vdots \\
		Q_{35}(\tau- \tau_{\mathrm{ep}}) & \dots & Q_{35}^{(l)}(\tau)
	\end{bmatrix}.
\end{equation}
The correlation matrix is calculated as
\begin{equation}
	C(\tau) = \frac{1}{| \tau_{\mathrm{ep}} |} Q(\tau) Q^{ \dagger}(\tau).
\end{equation}
This matrix contains as the element at position $i,j$ the Pearson correlation coefficient between $q_i (t)$ and $q_j (t)$ during epoch $\tau$. In case of missing values in the data, we disregard all values of that time stamp to ensure the calculated matrix is a positive-definite correlation matrix. We define a Euclidean distance measure between two matrices in the epochs $\tau$ and $\tau'$
\begin{equation}
	\label{eq:distancemeasure}
	d(\tau, \tau') = \sqrt{\sum_{i,j}(C_{ij}(\tau)-C_{ij}(\tau'))^2} = ||C^(\tau) - C^(\tau')|| ~.
\end{equation}
and apply $k$-means clustering \cite{Tan2019} with $k=2$ to all calculated matrices. This yields the cluster centers (element-wise mean matrices) shown in Fig. \ref{fig:clustercenters} and the distribution of the cluster appearances over daytime as shown in Fig. \ref{fig:clusterOverTime}. Clearly, cluster 1 shows times where the traffic flow at all cross-sections is strongly correlated, whereas cluster 2 is mostly uncorrelated. Here, only close to the diagonal some correlations remain, which indicate similar flows on neighboring detectors. Splitting further yields only a separation of the strongly correlated cluster based on the strength of the correlations, no structural differences are detected. Furthermore, we applied also hierarchical $k$-means clustering \cite{Munnix2012}, hierarchical clustering based on Ward's optimization criterion \cite{Ward1963, Murtagh2014}, complete-linkage and single-linkage clustering \cite{Legendre1998} to the matrices. We have tested all algorithms for solutions of two, three and four clusters. We studied the resulting cluster centers and also calculated the silhouette coefficient \cite{Rousseeuw1987} for each matrix
\begin{equation}
	s(\tau) = \left\{
	\begin{array}{ll}
		\displaystyle{\frac{b(\tau)-a(\tau)}{\max(b(\tau), a(\tau))}} & , ~|z_{n(\tau)}| > 1 \\
		0 & , ~|z_{n(\tau)}| = 1
	\end{array}
	\right. 
\end{equation}
where $n(\tau)$ gives the cluster corresponding to epoch $\tau$, and $|z_{n}|$ the elements in cluster $n$. $a(\tau)$ is the average distance to all other matrices in the same cluster
\begin{equation}
	a(\tau) = \frac{1}{|z_{n(\tau)}|-1} \sum_{\tau' \in z_{n(\tau)}, \tau' \neq \tau} d(\tau, \tau')
\end{equation}
and $b(\tau)$ the smallest average distance to a single other cluster
\begin{equation}
	b(\tau) = \min_{m \neq n(\tau)}\left( \frac{1}{|z_m|} \sum_{\tau' \in z_m} d(\tau, \tau') \right) ~.
\end{equation}
This coefficient takes values between $-1$ and $+1$ with larger positive values representing matrices that are well clustered and negative values showing matrices that are closer to another cluster than to their own. An indicator for the overall clustering is given by the average silhouette coefficient. This is shown in table \ref{tab:silcoefs} for all calculated clusterings. The hierarchical clustering with $k$-means splits and the one based on Ward's criterion yield very similar results to the standard $k$-means. This is not surprising as the optimization criteria for the algorithms are similar, but highlights that the found solution provides a reasonable grouping. The complete-linkage clustering gives a similar solution for two clusters, but starts to classify very small groups from then on. It is noteworthy that the two cluster solution has a higher silhouette coefficient than the $k$-means solution. However, the matrices and distribution over daytime show no significant differences for two clusters. As one would expect, the single-linkage clustering gives very different results, which do not separate the main groups as well, but rather lead to the detection of outliers. This is, of course, because the optimization criterion is very different from the others. We stick with the standard $k$-means solution for the following analysis. We have seen that its solution is reasonable and stable. It does not diverge towards outsider characterization for more than two clusters, which makes it a good first choice for this sort of classification also in other systems. In general, one has to tailor the identification of the non-stationarity to the problem at hand. We have found that correlation matrix clustering is useful in many systems \cite{Munnix2012, WangShanshan2020, Bette2022}.

A closer look at Fig. \ref{fig:clusterOverTime} allows interpretation of the clusters. The shown histogram counts reveal which cluster is dominant during which time of the day. The strongly correlated times in cluster 1 are caused by an overall increase (5am - 7am) and an overall decrease (8pm) in traffic volume. Another, less pronounced decrease is found during the 0am epoch. While the first two are caused by the major motion between low traffic flows during the night and high ones during the day, the latter is explained by traffic shifts from late evening traffic with low volumes to night time traffic with almost zero volume. As the 0am change is not as strong, cluster 2 remains dominant here. Generally, the appearance of cluster 1 marks transition periods between times with low and high traffic volume. Therefore, we will further split the two correlation clusters: Only consecutive times of a dominating correlation cluster are considered to be one cluster, the next appearance of the same correlation cluster is taken to be a new cluster. Thereby we achieve a good separation of times with different mean values, standard deviations and correlation structures. We arrive at five clusters:
\begin{enumerate}
	\item 0am - 4am,
	\item 5am - 7am,
	\item 8am - 7pm,
	\item 8pm,
	\item 9pm - 11pm.
\end{enumerate}
Of course, we could merge the first and last cluster, but for simplicity we leave the division as it is.
\begin{table}
	\caption{Silhouette coefficients for different clustering methods and solutions with 2, 3 and 4 clusters.}
	\centering
	\begin{tabular}{lllllll}
		\toprule
		method & 2 clusters & 3 clusters & 4 clusters \\
		\midrule
		$k$-means & 0.423 & 0.317 & 0.168 \\
		hierarchical $k$-means & 0.423 & 0.354 & 0.327 \\
		Ward's criterion & 0.379 & 0.330 & 0.168  \\
		complete-linkage & 0.451 & 0.452 & 0.345  \\
		single-linkage & 0.446 & 0.235& 0.219 \\
		\bottomrule
	\end{tabular}
	\label{tab:silcoefs}
\end{table}

Next, we project all the normal workday data points into the eigenvector systems of the corresponding cluster state analogously to the mean projection in equation \eqref{eq:projection}. This yields the reference normal distributions for the calculation of the Hellinger distance.

\begin{figure}
			\centering
		\includegraphics[width=0.7\textwidth]{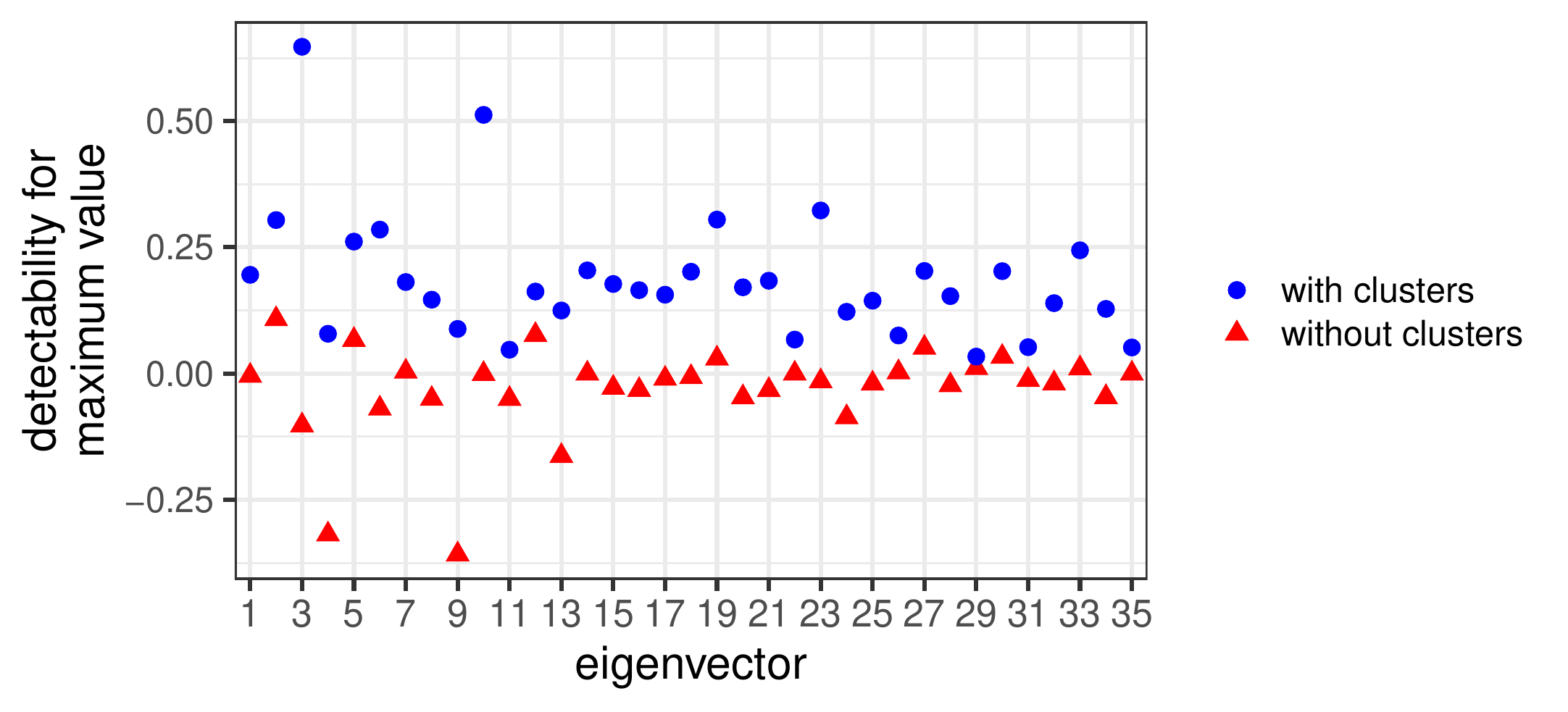}
		\caption{Detectability of a weekend day in case (a), where detectability is defined via a single exceedance of the detection threshold during the day. The detectability is shown for each eigenvector (principal component) of the system. Results are shown with knowledge about the non-stationarity during the day, i.e. with clusters, and without.}
		\label{subfig:exampleDetWeekendMax}
	\end{figure}

\begin{figure}
		\centering
		\includegraphics[width=0.7\textwidth]{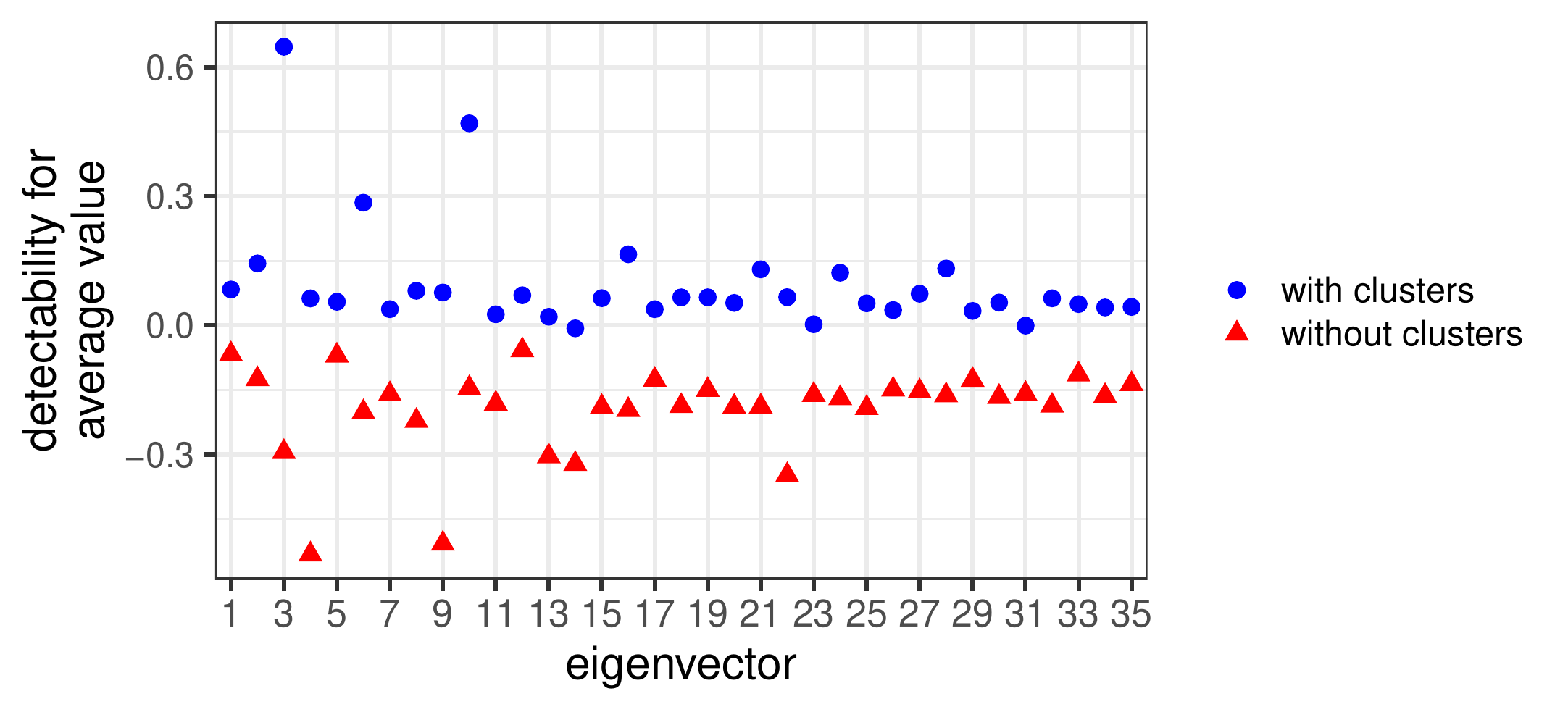}
		\caption{Detectability of a weekend day in case (b), where detectability is defined as an average exceedance of the threshold in one cluster. The detectability is shown for each eigenvector (principal component) of the system. Results are shown with knowledge about the non-stationarity during the day, i.e. with clusters, and without.}
		\label{subfig:exampleDetWeekendMean}
	\end{figure}

\begin{figure}
		\centering
		\includegraphics[width=0.7\textwidth]{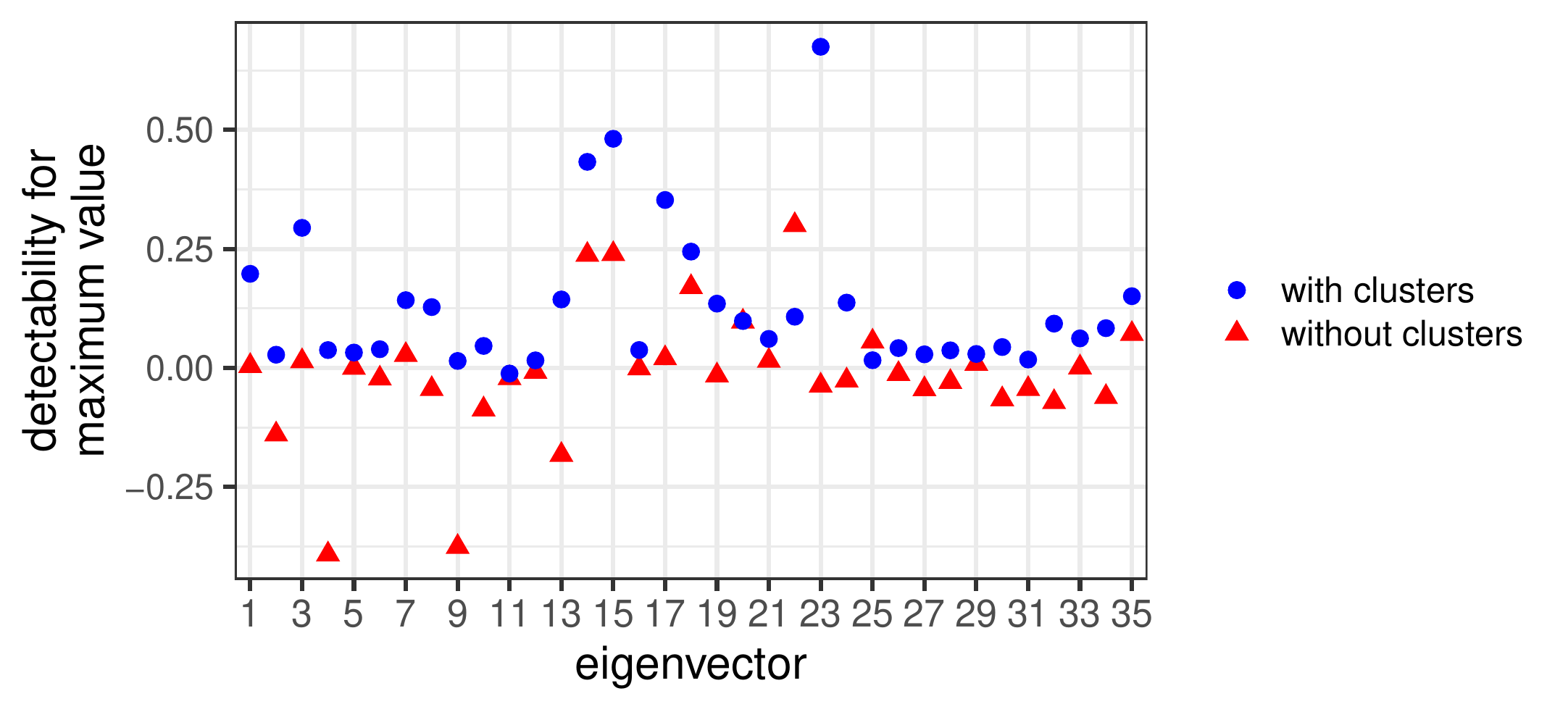}
		\caption{Detectability of the summer bank holiday onset in case (a), where detectability is defined via a single exceedance of the detection threshold during the day. The detectability is shown for each eigenvector (principal component) of the system. Results are shown with knowledge about the non-stationarity during the day, i.e. with clusters, and without.}
		\label{subfig:exampleDetHolidayMax}
	\end{figure}

\begin{figure}
		\centering
		\includegraphics[width=0.7\textwidth]{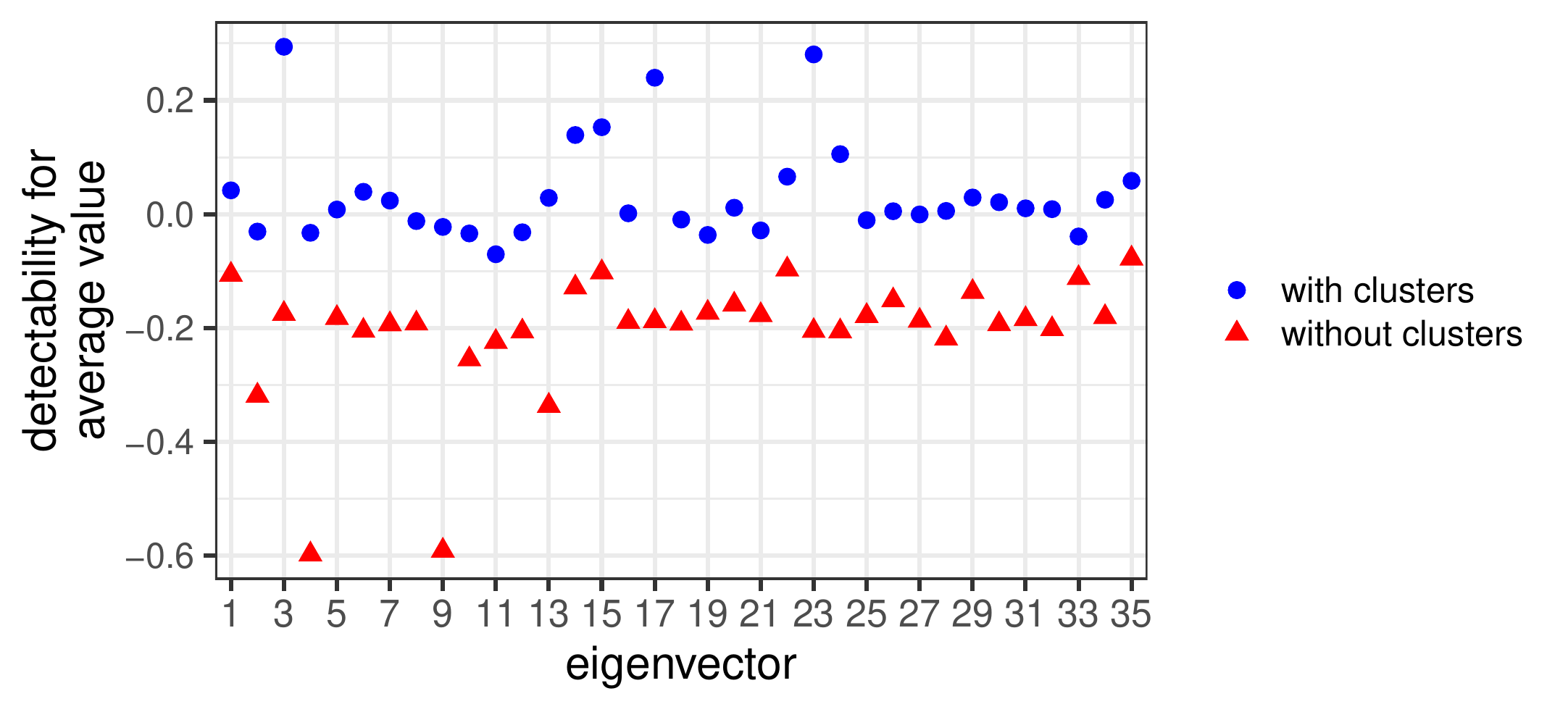}
		\caption{Detectability of the summer bank holiday onset in case (b), where detectability is defined as an average exceedance of the threshold in one cluster. The detectability is shown for each eigenvector (principal component) of the system. Results are shown with knowledge about the non-stationarity during the day, i.e. with clusters, and without.}
		\label{subfig:exampleDetHolidayMean}
	\end{figure}

We then choose a random normal workday and a random Sunday. The workday is needed to establish which Hellinger distances appear even without any changes due to fluctuations. Hence, each hour of the two days is projected into the corresponding eigenvectors and the resulting distributions are compared to the reference normal distributions by means of the Hellinger distance. Within each cluster $c$ we get a maximum appearing Hellinger distance for the normal data $H_{c}^{\mathrm{norm}}$. We define this as the threshold a test data point needs to exceed to be called a system change. This is a rather strict definition minimizing false positives for system changes or anomalies. For the weekend data we calculate $H_c^{\mathrm{anom}}$ once as the maximum (case (a)) and once as the mean (case (b)) over the daily Hellinger distances. Thus, we introduce the exceedance of the threshold
\begin{equation}
	d_c = H_{c}^{\mathrm{norm}} - H_c^{\mathrm{anom}}
\end{equation}
as a useful mesasure for detectability in cluster $c$. The overall detectability is then taken as the maximum over all clusters. This yields a detectability performance for each eigenvector. To determine if accounting for the non-stationarity is necessary, we perform the same analysis, but without clustering, i.e. there is only one cluster. The results for these calculations are shown in Figs. \ref{subfig:exampleDetWeekendMax} and \ref{subfig:exampleDetWeekendMean}. Detection is obviously easier with knowledge of the non-stationarity. Without, some eigenvectors are not at all usable for detection (values below zero). This is especially true for case (b). With clusters and a careful choice of the correct eigenvectors, we detect the change not only in a single data point but in the average performance of at least one cluster. In general, all eigenvectors show smaller detection performance without clusters. This comparison is not always perfectly fair as there is no warranty that eigenvector $j$ in the no cluster analysis corresponds to the same behavior as eigenvector $j$ in the cluster analysis. This was neglected in the simulation study in previous chapters as it does not matter in the Monte Carlo average. In Figs. \ref{subfig:exampleDetHolidayMax} and \ref{subfig:exampleDetHolidayMean} we show the same type of analysis once more, but instead of a random weekend day, we have taken the start of the summer bank holidays. Here, the mentioned fact is even more evident: For the system as a whole detectabilities are higher with clusters, but for single eigenvectors it might not be so. For example, in this case eigenvector 22 in the analysis with clusters and eigenvector 22 in the one without clusters do not have the same structure, i.e. they represent different behavior patterns of the multivariate system.

All together, we see a much better change detection performance in this example when the non-stationarity is accounted for. Especially, if we do not want to detect the change only in single data points (case (a)), but rather with a consistently high indicator (case (b)), it is impossible to detect without clusters. This is inferred from all results for the detectability without clusters showing negative values in case (b), cp. Figs. \ref{subfig:exampleDetWeekendMean} and \ref{subfig:exampleDetHolidayMean}. While of course in the case with clusters, less data points are used to calculate the mean per cluster, one can clearly identify the change in one of them. Without clusters there might be single points exceeding the detection threshold, but overall the Hellinger distances do not lie above the threshold.

The trend that changes are detected more easily in projections onto eigenvectors with small eigenvalues cannot be confirmed by a single example system. Which eigenvector is most suitable for detection depends on the interaction between correlation structure and system change. The aforementioned trend is therefore only true in the average over many systems.

\section{Conclusion}
\label{sec:conclusion}

We studied the sensitivity for change detection of principal components in non-stationary, correlated systems with multiple time series measurements. The non-stationarity was defined as the possibility for the system to be in multiple, distinctly different normal states prior to the change. Our study was based on the one conducted by Tveten \cite{Tveten2019} for stationary time series. Accordingly, the changes, which should be detected, were either to correlation structure, mean values or standard deviations of the time series. For simplicity, we constricted the study to those scenarios, where the normal states differ in the same property, in which the change also occurred. We analyzed how the detectability of the change varied for each principal component depending on the knowledge of the non-stationarity.

We found that in general the knowledge about the non-stationarity always increases the sensitivity for change detection. The increase is dependent on the principal component on which the data was projected. This dependency is quite similar to the one that was already found for the pure sensitivity for change detection in a stationary system. This means that where sensitivity for change is highest, also the increase gained by knowledge about the non-stationarity is highest. Usually this is for the minor components, i.e. the eigenvectors associated with small eigenvalues. This is reasonable as they represent behavior that is entirely unusual for a system in its normal state. When a change occurs, it can be seen very easily in those projections. If they are mixed up due to multiple normal states, this clear possibility to see changes is diminished. This underlines the importance of non-stationarity for direct uses of PCA for change detection and is confirmed in the traffic flow example.

Usage of PCA for dimensionality reduction by keeping only projections on major components will probably be less influenced by this. This is an interesting aspect to study in the future. Other methods depending on eigenvectors for change detection (e.g. Mahalanobis distance) were not directly studied, but we think it likely that their sensitivity to change could also be increased by the consideration of multiple normal states. Our results are true for all three different types of changes and states. We further analyzed two scenarios: Unrelated and related normal states. We found that the sensitivity increase is usually greater for unrelated states, which is reasonable as the non-stationarity has a stronger effect in this case. We want to point out once again, that the use of multiple normal states for change detection in real applications is only possible, if a criterion can be found to identify which normal states new data should be compared to. The purpose of the present study was mainly to develop the concepts and to provide the necessary tools. The traffic flow data example shows that it is possible to transfer the idea onto real world data. This opens up applicability in many systems, where PCA and related methods are used for change detection. While traffic is among these systems \cite{Li2019}, other prominent examples for fault detection are chemical plants \cite{Nawaz2021} and industrial machinery \cite{Miguel2022,Pozo2018}. A first step to include multiple operational conditions in PCA-based failure detection for heat pumps was already undertaken by Zhang et. al. \cite{Zhang2019}. We intend to test the proposed method on wind turbines, where multiple operational normal conditions based on correlation matrix clustering have already been found \cite{Bette2022} and PCA is being used for fault detection \cite{Vidal2018}. 

Our results clearly show that non-stationarity should be taken into account if one undertakes change, novelty or failure detection using principle component analysis.

\section*{Acknowledgements}
We acknowledge fruitful conversations with Joachim Peinke, Matthias Wächter, Christian Phillipp, Timo Lichtenstein and Anton Heckens. This study was carried out in the project \textit{Wind farm virtual Site Assistant for O\&M decision support – advanced methods for big data analysis} (WiSAbigdata) funded by the Federal Ministry of Economics Affairs and Energy, Germany (BMWi). One of us (H.M.B.) thanks for financial support in this project.


\printbibliography

\end{document}